\documentclass[prd,twocolumn,showpacs,superscriptaddress,eqsecnum,nofootinbib,floatfix]{revtex4}

\usepackage{amsfonts}
\usepackage{amsmath}
\usepackage{amssymb}
\usepackage{amsthm}
\usepackage{bm}
\usepackage{dcolumn}
\usepackage{epsfig}
\usepackage{graphicx}
\usepackage{graphics}
\usepackage{latexsym}
\usepackage{rotating}
\usepackage[dvips]{color}

\newcommand\speculate{\buildrel {?}\over{=}}

\begin{document}
\title{Resonantly enhanced and diminished strong-field
  gravitational-wave fluxes}

\author{\'Eanna E.\ Flanagan}
\affiliation{Center for Radiophysics and Space Research, Cornell
  University, Ithaca, NY 14853, USA}

\author{Scott A.~Hughes}
\affiliation{Department of Physics and MIT Kavli Institute, MIT,
  Cambridge, MA 02139, USA}
\affiliation{Canadian Institute for Theoretical Astrophysics,
  University of Toronto, 60 St.\ George St., Toronto, ON M5S 3H8,
  Canada}
\affiliation{Perimeter Institute for Theoretical Physics, Waterloo, ON
  N2L 2Y5, Canada}

\author{Uchupol Ruangsri}
\affiliation{Department of Physics and MIT Kavli Institute, MIT,
  Cambridge, MA 02139, USA}

\date{\today}

\begin{abstract}
The inspiral of a stellar mass ($1 - 100\,M_\odot$) compact body into
a massive ($10^5 - 10^7\,M_\odot$) black hole has been a focus of much
effort, both for the promise of such systems as astrophysical sources
of gravitational waves, and because they are a clean limit of the
general relativistic two-body problem.  Our understanding of this
problem has advanced significantly in recent years, with much progress
in modeling the ``self force'' arising from the small body's
interaction with its own spacetime deformation.  Recent work has shown
that this self interaction is especially interesting when the
frequencies associated with the orbit's $\theta$ and $r$ motions are
in an integer ratio: $\Omega_\theta/\Omega_r = \beta_\theta/\beta_r$,
with $\beta_\theta$ and $\beta_r$ both integers.  In this paper, we
show that key aspects of the self interaction for such ``resonant''
orbits can be understood with a relatively simple
Teukolsky-equation-based calculation of gravitational-wave fluxes.  We
show that fluxes from resonant orbits depend on the relative phase of
radial and angular motions.  The purpose of this paper is to
illustrate in simple terms how this phase dependence arises using
tools that are good for strong-field orbits, and to present a first
study of how strongly the fluxes vary as a function of this phase and
other orbital parameters.  Future work will use the full dissipative
self force to examine resonant and near resonant strong-field effects
in greater depth, which will be needed to characterize how a binary
evolves through orbital resonances.
\end{abstract}


\maketitle

\section{Introduction}
\label{sec:intro}

\subsection{The self-force driven evolution of binaries:\\
A very brief synopsis}
\label{sec:synopsis}

Our understanding of the two-body problem in general relativity has
advanced substantially in the past decade.  Besides the celebrated
breakthroughs in numerical relativity {\cite{pretorius05,
    campanelli06, baker06}} which have opened the field of binary
phenomenology in general relativity, there has been great progress in
understanding the extreme mass-ratio limit of this problem, when one
member of the binary is much smaller than the other.  This limit is of
great interest in describing astrophysical extreme mass-ratio binaries
(a particularly interesting source for space-based gravitational-wave
measurements) {\cite{emri}}, and as a limiting form of the more
generic two-body problem {\cite{lousto,sperhake}}.

Most efforts to model extreme mass-ratio binaries have focused on the
computation of {\it self forces} (see Ref.\ {\cite{barack}} for a
recent comprehensive review).  Consider a small body orbiting a black
hole.  At zeroth order in the small body's mass, its motion is
described as a geodesic of the black hole spacetime.  At first order
in this mass, the black hole's spacetime is slightly deformed.  This
deformation changes the trajectory that the small body follows,
pushing it away from the background spacetime's geodesic.  It is
useful to regard the change to the trajectory as arising from a self
force which modifies the geodesic equations typically used to describe
black hole orbits.  Conceptually, it is useful to split the self force
into two pieces: A time-symmetric {\it conservative} piece, and a
time-asymmetric {\it dissipative} piece.  On average, the impact of
the conservative contribution is to shift orbital frequencies away
from their geodesic values.  The dissipative self force is equivalent,
on average, to a slow evolution of the otherwise conserved constants
(e.g., the orbital energy and angular momentum) which characterize
geodesic orbits.  It makes the largest contribution to an orbit's
phase evolution.  The conservative piece makes a smaller (though still
significant) contribution which accumulates secularly over many orbits
{\cite{ppn,pp}}.

Recent work by Flanagan and Hinderer {\cite{FH}} (hereafter FH) using
a post-Newtonian (pN) approximation to the self force together with
fully relativistic orbital dynamics has shown that a small body's self
interaction becomes particularly important near {\it resonances}.  The
background geodesic motion can be characterized by three orbital
frequencies with respect to Boyer-Lindquist time: A radial frequency
$\Omega_r$, a polar frequency $\Omega_\theta$, and an axial frequency
$\Omega_\phi$.  In the weak-field (large separation) limit, these
three frequencies asymptote to the Newtonian Kepler frequency.  In the
strong field, these frequencies can differ significantly, with
$\Omega_r$ always the smallest frequency (the relative magnitude of
$\Omega_\theta$ and $|\Omega_\phi|$ depends on the sign of the orbit's
axial angular momentum). Resonant orbits are ones for which the radial
and angular motions become commensurate: $\Omega_\theta/\Omega_r =
\beta_\theta/\beta_r$, where $\beta_\theta$ and $\beta_r$ are small
integers with no common factors.  On such orbits, components of the
self interaction which normally ``average away'' when examined over a
full orbital period instead combine coherently, substantially changing
their impact on the system's evolution.

For the purpose of our background discussion, it is useful to include
more details from FH's analysis of how resonant effects arise.
Consider a body of mass $\mu$ moving on a bound trajectory near a Kerr
black hole of mass $M$, with $\mu \ll M$.  FH note that one can
describe the motion of this body using action-angle variables and
correctly accounting for how the integrals which parameterize geodesic
orbits evolve due to the self force.  Writing the angle variables
$q_\alpha = (q_t,q_r,q_\theta,q_\phi)$ (which describe motions in the
$t$, $r$, $\theta$, and $\phi$ directions of Boyer-Lindquist
coordinates), and writing the integrals associated with geodesic
motion $J_i = (E, L_z, Q)$ (with $E$ the energy, $L_z$ the axial
angular momentum, and $Q$ the Carter constant), the equations of
motion describing the system are {\cite{FHtt}}
\begin{eqnarray}
\frac{dq_\alpha}{d\tau} &=& \omega_\alpha({\bf J}) + \epsilon
g^{(1)}_\alpha(q_r,q_\theta,{\bf J}) + O(\epsilon^2)\;,
\label{eq:eom1}\\
\frac{dJ_i}{d\tau} &=& \epsilon G^{(1)}_i(q_r,q_\theta,{\bf J})
+ O(\epsilon^2)\;.
\label{eq:eom2}
\end{eqnarray}
The time parameter $\tau$ is proper time along the orbit; the
parameter $\epsilon = \mu/M$, the system's mass ratio.  The
$\omega_{r,\theta,\phi}$ are fundamental frequencies with respect to
proper time associated with bound Kerr geodesic orbits.  The forcing
functions $g^{(1)}_\alpha$ and $G^{(1)}_i$ arise from the first-order
self force.  FH also include discussion of second-order forcing
functions, which we do not need for this synopsis; see
Ref.\ {\cite{FH}} for further discussion.

At order $\epsilon^0$, Eqs.\ (\ref{eq:eom1}) and (\ref{eq:eom2})
simply describe geodesics of Kerr black holes: The integrals of the
motion are constant, and each angle variable evolves according to its
associated frequency.  The leading adiabatic dissipative correction to
this motion can be found by dropping the forcing term $g^{(1)}_\alpha$
and replacing $G^{(1)}_i$ by $\langle G^{(1)}_i\rangle$, the average
of this forcing term over the 2-torus parameterized by $q_\theta$ and
$q_r$ {\cite{FHtt}}.  To compute this torus-averaged self force, it is
sufficient to use the radiative approximation {\cite{mino03,FHtt,pp}},
which includes only the radiative contributions to the self
interaction and neglects conservative contributions.  For generic
(non-resonant) orbits, this torus average coincides with an infinite
time average, and the averaged quantities $\langle G^{(1)}_i \rangle$
are just the time-averaged fluxes of energy, angular momentum and
Carter constant.  In recent years such time-averaged fluxes have been
computed numerically using the frequency domain Teukolsky equation
{\cite{dh06,fujita,sago}}.  These fluxes can be used to compute
leading-order, adiabatic inspirals.  The conservative contributions
influence the motion only beyond the leading adiabatic order
{\cite{mino03,FHtt}}.

\subsection{Resonant effects}
\label{sec:res_intro}

Now consider going beyond the leading adiabatic order.  Important
post-adiabatic effects can be found by continuing to neglect
$g^{(1)}_\alpha$, but now integrating Eq.\ (\ref{eq:eom2}) using
$G^{(1)}_i$ rather than its averaged variant.  FH show that for
``most'' orbits, $G^{(1)}_i$ is given by $\langle G^{(1)}_i\rangle$
plus a rapidly oscillating contribution.  Over the timescales
associated with inspiral, this rapidly oscillating piece averages away
and has little effect.  The effect of the forcing term $G^{(1)}_i$ is
dominated by $\langle G^{(1)}_i\rangle$ for all non-resonant orbits.

For resonant orbits, this averaging fails: contributions beyond
$\langle G^{(1)}_i\rangle$ are {\it not} rapidly oscillating, and can
significantly modify how the integrals of motion evolve during an
inspiral.  A given binary is very likely to evolve through several
low-order resonances en route to the final merger of the smaller body
with the large black hole {\cite{rh_inprep}}.  A complete quantitative
understanding of these resonant effects will thus be quite important
for making accurate inspiral models.  Prior to FH's analysis, several
other papers argued that such resonances may play an important role in
the radiative evolution of binary systems {\cite{mino05,tanaka06}}
(albeit without quantifying the detailed impact they can have), or
else because of other effects which resonances have on the evolution
of a dynamical system {\cite{acl10}}.

Orbits in which $\Omega_\theta/\Omega_r$ take on a small-integer ratio
have been studied in great detail by Grossman, Levin, and Perez-Giz
{\cite{glpgI}}, who called them ``periodic'' orbits and provided a
fairly simple scheme for classifying their features.  Following
Ref.\ {\cite{FH}} (as well as more recent work by Grossman, Levin, and
Perez-Giz {\cite{glpgII}}), we will call them ``resonant'' orbits,
reflecting the fact that our main interest is in understanding how
their periodic structure impacts the self interaction.  Grossman,
Levin and Perez-Giz have more recently argued for the utility of using
resonant orbits as sample points in numerical computations of leading
order, adiabatic inspirals: evaluating fluxes at resonant orbits may
enable a speedup of flux computations {\cite{glpgII}}, more
efficiently covering the parameter space of generic orbits.  Although
their goals are rather different from ours here, many of their
techniques and results substantially overlap with ours (modulo minor
differences in notation).  We highlight the overlap at appropriate
points in this paper.

As a binary evolves through a resonance, its self interaction and thus
its evolution are modified compared to what we would expect if the
resonance were not taken into account.  The details of how the self
interaction is modified depend on the relative phase of the radial and
angular motions as the orbit passes through resonance.  Because of
this, {\it resonances enhance the dependence of a binary's orbital
  evolution on initial conditions.}  Let the phase variable $\chi_0$
define the value of the orbit's $\theta$ angle at the moment it
reaches periapsis (see Sec.\ {\ref{sec:generalgeod}} for more
details).  On resonance, two orbits which have the same energy $E$,
the same axial angular momentum $L_z$, and the same Carter constant
$Q$ will evolve differently if they have different values of $\chi_0$.

FH estimate {\cite{FH}} that the shift to the orbital phase induced by
these resonances can be several tens to $\sim 10^2$ radians for mass
ratios $\sim 10^{-6}$ (as compared to an analysis which neglects the
resonances).  That there is such a large shift, and that this shift
may depend on initial conditions, is potentially worrisome.
Resonances could significantly complicate our ability to construct
models for measuring the waves from extreme mass-ratio inspirals.  On
the other hand, the detailed behavior of a system as it evolves
through resonances may offer an opportunity to study an interesting
aspect of strong-field gravity, providing a new handle for
strong-gravity phenomenology.  Analytic studies of the effect of the
passage through resonance can be found in Refs.\ \cite{gyb11,fh13}.

\subsection{Our analysis}
\label{sec:our_paper}

The ``several tens to $\sim 10^2$ radians'' estimate by FH is based on
applying pN self force estimates to strong-field orbits, a regime
where pN approximations are generally inaccurate.  It is thus of great
interest to estimate the impact of orbital resonances using
strong-field methods.  The purpose of this paper is to take a first
step in this direction.

Our goal is to generalize our computational techniques in order to
treat resonances correctly.  A key point is that the flux-balancing
technique which can be used to approximate inspiral (as described in
the final paragraph of Sec.\ {\ref{sec:synopsis}}) is based on the
adiabatic approximation.  This approximation temporarily breaks down
during a resonance.  Therefore, to treat resonances, one must use the
orbital equations of motion (\ref{eq:eom1}) -- (\ref{eq:eom2}) to
track the evolution of all the orbital degrees of freedom on short
timescales.  Flux balancing instead just tracks the evolution of the
conserved quantities $E$, $L_z$ and $Q$ on long timescales.  In
addition one must use the full, oscillatory self-force driving term
$G^{(1)}_i$, and not just its averaged version.

As is well known, computation of the full self force is extremely
difficult, largely because it requires regularization of the self
field \cite{barack}.  Fortunately, only the dissipative piece of the
self force should contribute to leading order resonance effects.  As
argued in FH, there is some evidence suggesting that geodesic motion
perturbed by the conservative piece of the self force is an integrable
dynamical system, and resonances do not occur in such systems.  Thus,
if the integrability conjecture of FH is true, only the dissipative
self force needs to be computed.  This constitutes a great
simplification, since the well-known difficulties of self-force
computations apply only to the conservative piece; the dissipative
piece is relatively straightforward to compute.  Techniques for doing
so with scalar fields were presented in Ref.\ {\cite{dfh05}}, and
generalizing to the gravitational dissipative self force is not
terribly difficult {\cite{mino05,tanaka06}}.  While these references
focused on the {\it averaged} self force, it is straightforward to
generalize the analysis to obtain the full dissipative self force.

It is thus feasible to perform numerical compututations of orbital
evolutions through resonances using the full dissipative self force,
without any orbit averaging.  Our eventual goal is to extend our black
hole perturbation theory codes to do just this, and to evaluate how
the dissipative self force behaves as a system evolves through
resonance.  Work in this vein is in progress , and will be presented
in future work {\cite{fhhr_inprep}}.

In this paper, we take a first step in this direction.  We focus here
on computation of time-averaged fluxes of the integrals of the motion,
and in particular on how these quantities differ between resonant and
non-resonant orbits.  These quantities correspond to the fluxes that
one would measure at infinity (and at the black hole horizon) if one
turned off radation reaction effects; upon averaging over long times,
they are equal to the rate at which the dissipative self force evolves
these constants.  We emphasize that these quantities are not
sufficient to allow computation of orbital evolutions.  However, they
provide insight into the characteristic features of the radiation
emitted by resonant orbits.

We find that fluxes from resonant orbits generically differ from those
from nearby, non-resonant orbits\footnote{Thus the fluxes change
  discontinuously as one varies the orbital parameters.  This is
  certainly unphysical, but arises because we compute infinite time
  averages of fluxes from geodesic orbits.  If one considers the
  fluxes from the true inspiraling motion, and averages over a
  timescale intermediate between the orbital timescale and the
  radiation reaction timescale, the time-averaged fluxes would vary
  smoothly with time, with order unity changes in the vicinity of
  resonances.  This point is discussed further in Appendix
  {\ref{app:qdot}}.}, and in addition vary depending on the relative
phase of the radial and angular motions.  The magnitude of this
variation is closely related to the ``kick'' that is imparted to the
orbit's constants as it evolves through a resonance (cf.\ Fig.\ 1 of
FH).  As such, characterizing on-resonance fluxes is a useful and
natural first step in the process of modifying existing flux-based
codes to compute the full dissipative self force.  We explore
numerically the magnitude of the difference between the resonant and
non-resonant cases, and the dependence on the orbital phase.  For
specific modes, the fluxes can vary by large factors (although
variations of order unity are more typical).  For the net fluxes
obtained by summing over all modes, variations are typically of order
a percent or less.

\subsection{Outline of this paper}
\label{sec:outline}

We begin this paper by briefly reviewing the behavior of Kerr geodesic
orbits in Sec.\ {\ref{sec:geodesics}}.  Much of this material has been
presented elsewhere, so we leave out most details, pointing the reader
to appropriate references.  Our main focus is to describe how to find
and characterize resonant orbits.  We then describe how to compute
radiation from Kerr orbits in Sec.\ {\ref{sec:radiate}}.  We first
briefly review the Teukolsky-equation-based formalism we use
(Secs.\ {\ref{sec:fdteuk}} -- {\ref{sec:fluxes}}), and then describe
how key details are modified by orbital resonances in
Secs.\ {\ref{sec:resI}} and {\ref{sec:resII}}.  We describe two
complementary approaches to computing fluxes on resonance.  Although
formally equivalent (as we prove in Appendix {\ref{app:equivalence}}),
their implementation is quite different.  Having both methods at hand
proved useful to us in our numerical study.  One aspect of the
on-resonance computation (the evolution of Carter's constant $Q$) is
sufficiently complicated that all details of this calculation are
given in Appendix {\ref{app:qdot}}.  Our analytic results for fluxes
of energy and angular momentum on resonance agree with those obtained
by Grossman, Levin and Perez-Giz (compare especially Secs.\ IIID--E
and Appendices B5, B6, and C in Ref.\ {\cite{glpgII}} with our
discussion here, and with our Appendix {\ref{app:equivalence}}).  Our
result for the resonant rate of change of the Carter constant appears
to be new.

Our numerical results are given in Sec.\ {\ref{sec:results}}.  We
begin by examining how fluxes from individual modes (i.e., harmonics
of the orbital frequencies) behave as a function of the offset phase
of the radial and angular motions, which we denote $\chi_0$.  We show
that the amplitude of a given mode, and hence the rates of change of
conserved quantities associated with that mode, can vary significantly
with $\chi_0$.  For example, the flux of energy from an orbit can vary
by factors of order unity as $\chi_0$ varies from $0$ to $2\pi$.  The
rate of change of the orbit's Carter constant can even change sign as
$\chi_0$ varies.  The total flux from a given orbit is given, however,
by adding fluxes from many modes.  When many modes are combined, much
of the variation washes away; we find variations of a fraction of a
percent in most quantities after summation.  The amount of this
residual variation seems to depend most strongly upon the geometry of
the orbit's $(r,\theta)$ motion on resonance, in particular the
topology of the trace in the $(r,\theta)$ plane.  Orbits whose motion
in $(r,\theta)$ have a simple topology with few trajectory crossings
in the plane (e.g., the $\Omega_\theta/\Omega_r = 3/2$ resonance) tend
to have relatively large variation in the integrals of motion; orbits
whose motion has a more complicated topology with many trajectory
crossings show much less variation (e.g., the $\Omega_\theta/\Omega_r
= 4/3$ resonance).  We argue that this can be explained in terms of
how the orbital motion tends (or fails) to average away variations in
the source-term to the Teukolsky equation.

As emphasized in Sec.\ {\ref{sec:our_paper}}, understanding these
fluxes exactly on resonance is only the first step in building a
complete strong-field understanding of how resonances impact
inspirals.  In particular, these results do not provide enough
information to specify how a system will evolve through a resonance.
To go further, it will be necessary to examine how dissipation behaves
as the system evolves toward and through an orbital resonance.  As
mentioned above, this analysis is now beginning; we briefly outline
the approach we are pursuing in Sec.\ {\ref{sec:conclude}}.

Throughout this paper, we use ``relativist's units,'' setting $G = 1 =
c$.

\section{Kerr geodesics and orbital resonances}
\label{sec:geodesics}

\subsection{Brief summary of general characteristics}
\label{sec:generalgeod}

We begin by reviewing geodesic orbits of Kerr black holes, with a
focus on aspects of this motion particularly relevant to our analysis.
In most textbooks [for example, Ref.\ {\cite{mtw}},
  Eqs.\ (33.32a)--(33.32d)], Kerr geodesics for a massive body are
described using equations of motion in the Boyer-Lindquist coordinates
$t$, $r$, $\theta$, and $\phi$:
\begin{eqnarray}
\Sigma^2\left(\frac{dr}{d\tau}\right)^2 &=& \left[E(r^2+a^2) - a
  L_z\right]^2
\nonumber\\
& & - \Delta\left[r^2 + (L_z - a E)^2 + Q\right]
\nonumber\\
&\equiv& R(r)\;,\label{eq:rdot}\\
\Sigma^2\left(\frac{d\theta}{d\tau}\right)^2 &=& Q - \cot^2\theta L_z^2
 -a^2\cos^2\theta(1 - E^2)\nonumber\\
&\equiv&\Theta(\theta)\;,\label{eq:thetadot}\\
\Sigma\left(\frac{d\phi}{d\tau}\right) &=&
\csc^2\theta L_z + aE\left(\frac{r^2+a^2}{\Delta} - 1\right)
-\frac{a^2L_z}{\Delta}\nonumber\\
&\equiv&\Phi(r,\theta)\;,\label{eq:phidot}\\
\Sigma\left(\frac{dt}{d\tau}\right) &=&
E\left[\frac{(r^2+a^2)^2}{\Delta} - a^2\sin^2\theta\right]
\nonumber\\
& & + aL_z\left(1 - \frac{r^2+a^2}{\Delta}\right)
\nonumber\\
&\equiv& T(r,\theta)\;.\label{eq:tdot}
\end{eqnarray}
In these equations, $\tau$ is proper time along the geodesic, $\Sigma
= r^2 + a^2\cos^2\theta$, and $\Delta = r^2 - 2Mr + a^2$.  The
quantities $E$ and $L_z$ are the orbital energy and axial angular
momentum, normalized to the mass $\mu$ of the orbiting body, and $Q$
is the orbit's Carter constant, normalized to $\mu^2$.  These three
quantities are conserved on any geodesic.

Along with the coordinate time $t$ and proper time $\tau$, it is often
very useful to work using a time parameter $\lambda$, defined by
$d\lambda = d\tau/\Sigma$.  The geodesic equations parameterized in
this way are
\begin{eqnarray}
\left(\frac{dr}{d\lambda}\right)^2 = R(r)\;,
&\qquad&
\left(\frac{d\theta}{d\lambda}\right)^2 = \Theta(\theta)\;,
\nonumber\\
\frac{d\phi}{d\lambda} = \Phi(r,\theta)\;,
&\qquad&
\frac{dt}{d\lambda} = T(r,\theta)\;.
\end{eqnarray}
By using $\lambda$ as our orbital parameter, the $r$ and $\theta$
coordinate motions are completely separated from one another.  Proper
time $\tau$ couples $r$ and $\theta$ by the factor $\Sigma$; the
coupling with coordinate time $t$ is even more complicated.  The
parameter $\lambda$ is often called ``Mino time,'' following Mino's
use of it to untangle these coordinate motions {\cite{mino03}}.

We have found it useful for many of our studies to introduce the
following reparameterization of $r$ and $\theta$:
\begin{equation}
r = \frac{pM}{1 + e\cos\psi}\;,\qquad
\cos\theta = \cos\theta_m\cos(\chi + \chi_0)\;.
\label{eq:rdefthdef}
\end{equation}
These transformations replace the variables $r$ and $\theta$ with
secularly accumulating angles $\psi$ and $\chi$.  As $\psi$ and $\chi$
evolve from 0 to $2\pi$, $r$ and $\theta$ move through their full
ranges of motion.  We define $\chi = \psi = 0$ at $\lambda = 0$.

Notice that we include an offset phase $\chi_0$ for the angular
motion.  We could also include an offset phase $\psi_0$ for the radial
motion, as well as initial conditions $\phi_0$ and $t_0$ for the
$\phi$ and $t$ coordinates.  We choose our time origin such that $t =
0$ when $\lambda = 0$, which means $t_0 = 0$.  We likewise choose
$\phi_0 = 0$.  Changing $\phi_0$ is equivalent to rotating around the
black hole's spin axis, and can have no effect on the flux of energy
and angular momentum from the system (although it introduces a phase
offset to the system's gravitational waves).

Finally, we choose $\psi_0 = 0$, which amounts to setting $\lambda =
0$ at a moment that the orbit passes through periapsis, $r = r_{\rm
  peri} = pM/(1 + e)$.  The offset phase $\chi_0$ thus sets the value
of $\theta$ at periapsis.  Previous work (e.g., {\cite{dh06}}) has
typically used $\chi_0 = 0$ as well.  The parameter set $(\psi_0,
\chi_0, \phi_0, t_0)$ is equivalent to the set $(\lambda^r_0,
\lambda^\theta_0, \phi_0, t_0)$ used in Ref.\ {\cite{dfh05}}.
Following this reference, $\chi_0 = 0$ will label the ``fiducial
geodesic.''  We will use it as a reference geodesic for some of the
calculations in Sec.\ {\ref{sec:radiate}}.

In their original form, Eqs.\ (\ref{eq:rdot}) -- (\ref{eq:tdot}), Kerr
orbits are parameterized (up to initial conditions) by the three
conserved constants $E$, $L_z$, and $Q$.  The reparameterization
(\ref{eq:rdefthdef}) maps those constants to parameters that describes
an orbit's coordinate geometry: semi-latus rectum $p$, eccentricity
$e$, and minimum angle $\theta_m$.  These quantities are likewise
conserved along a geodesic.  Schmidt {\cite{schmidt}} provides
closed-form expressions for converting between $(E,L_z,Q)$ and
$(p,e,\theta_m)$.  Either the set $(E,L_z,Q)$ or $(p,e,\theta_m)$,
plus the relative phase $\chi_0$, completely specifies a geodesic for
our purposes here.

\subsection{Orbital frequencies and resonances}
\label{sec:resonances}

Each orbit has a set\footnote{Interestingly, this set is {\it not}
  unique: There exists in the strong field geometrically distinct
  orbits (i.e., with different parameters $p, e, \theta_m$) that have
  identical frequencies.  See Ref.\ \cite{degenerate} for detailed
  discussion.} of frequencies describing its motions with respect to
$r$, $\theta$, and $\phi$.  The frequencies
\begin{equation}
\Omega_{r,\theta,\phi} = 2\pi/T_{r,\theta,\phi}
\end{equation}
are conjugate to the periods\footnote{Describing the periods using
  Boyer-Lindquist time $t$ is a bit complicated; $T_{r,\theta,\phi}$
  really describe an averaged notion of the periods.  See
  Refs.\ {\cite{schmidt,dh04}} for more detailed discussion.}
expressed in coordinate time $t$; the frequencies
\begin{equation}
\Upsilon_{r,\theta,\phi} = 2\pi/\Lambda_{r,\theta,\phi}
\end{equation}
are conjugate to these periods in Mino time $\lambda$.  These two
frequencies are related by a factor $\Gamma$ which describes the
average increase in $t$ per unit $\lambda$:
\begin{equation}
\Omega_{r,\theta,\phi} = \Upsilon_{r,\theta,\phi}/\Gamma\;.
\end{equation}
Details of how to compute these frequencies given $(E,L_z,Q)$ or
$(p,e,\theta_m)$ are given in Ref.\ {\cite{dh04,fh09}}.  One could
also define frequencies conjugate to proper time $\tau$ (see, e.g.,
Ref.\ {\cite{schmidt}} and discussion in Sec.\ {\ref{sec:intro}}), but
the $\Omega$ and $\Upsilon$ frequencies are sufficient for our
purposes.

\begin{figure*}[ht]
\includegraphics[width = 0.48\textwidth]{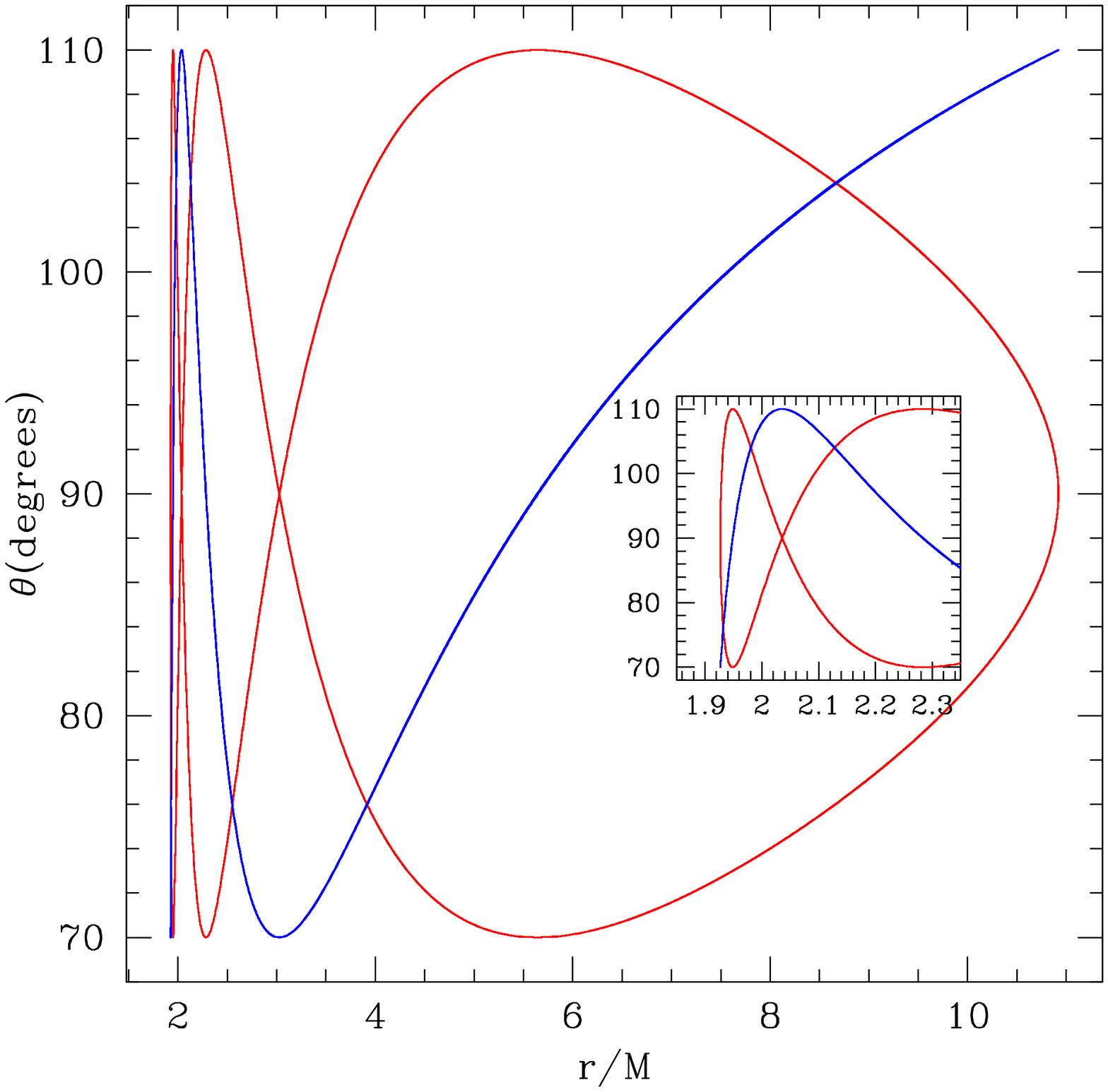}
\includegraphics[width = 0.48\textwidth]{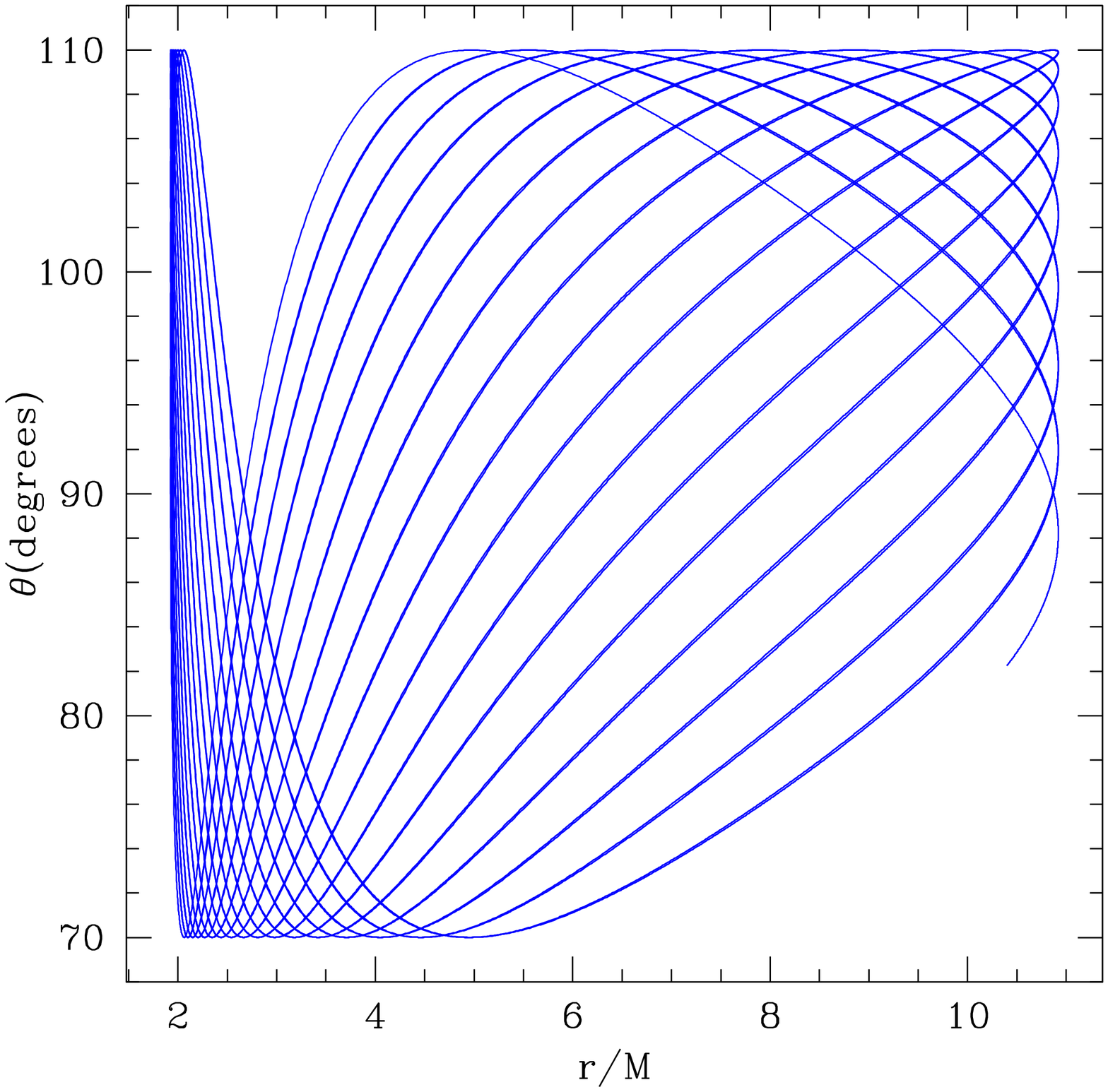}
\caption{Left: Lissajous figures describing motion in the $(r,\theta)$
  plane on a 3:1 orbital resonance ($a = 0.9M$, $p = 3.2758M$, $e =
  0.7$, $\theta_m = 70^\circ$).  The blue trace has $\theta =
  \theta_m$ at periapsis; red has $\theta = \pi/2$ at periapsis.  The
  inset image zooms in on the region $1.9 M \lesssim r \lesssim 2.3M$,
  clarifying the angular oscillation at very small radius.
  Approximately nine radial cycles are used to generate these traces.
  Right: Ergodic motion of a ``normal'' orbit.  The orbit's geometry
  is identical to that in the left-hand panel, but we have changed the
  black hole's spin to $a = 0.95M$; this changes the ratio of
  frequencies to $\Omega_\theta/\Omega_r = 2.0311\ldots$.  Again,
  roughly nine radial cycles are shown here.  Given enough time, this
  trace would pass arbitrarily close to all points in $70^\circ \le
  \theta \le 110^\circ$, $2M \alt r \alt 12M$.}
\label{fig:rtheta}
\end{figure*}

We next review how the qualitative features of the resonant orbits
differ from those of generic orbits, as background to
Sec.\ {\ref{sec:radiate}} below.  A more detailed discussion can be
found in Sec.\ II of Ref.\ {\cite{glpgII}}.  As an example, we compare
a typical orbit, for which the ratio $\Omega_\theta/\Omega_r$ is some
irrational number, to a resonant orbit, for which
$\Omega_\theta/\Omega_r = \beta_\theta/\beta_r$, where $\beta_\theta$
and $\beta_r$ are small integers with no common factors.  Figure
\ref{fig:rtheta} shows the motion of three orbits, projected into the
($r,\theta$) plane.  In all cases, we have chosen $p = 3.2758$, $e =
0.7$, $\theta_m = 70^\circ$; the motion is thus bound to the range
$1.93M \le r \le 10.9M$, $70^\circ \le \theta \le 110^\circ$.  (See
also Fig.\ 1 of Ref.\ {\cite{glpgII}}, which is very similar, although
it does not illustrate the impact of the offset phase between the
$r$ and $\theta$ motions.)

In the right-hand panel, we have set the spin parameter $a = 0.95M$.
For these orbital parameters, this orbit has $\Omega_\theta/\Omega_r =
2.0311\ldots$.  This is {\it not} a resonant orbit; notice that the
roughly nine radial periods shown here do not close.  The orbital
trace in this case ergodically fills the ($r,\theta$) plane.  In the
left-hand panel, we have set $a = 0.9M$, which yields
$\Omega_\theta/\Omega_r = 3$ --- these orbits are in a 3:1 resonance.
The two traces shown in this panel correspond to different choices of
$\chi_0$.  The blue trace has $\chi_0 = 0$ (so that $\theta =
\theta_m = 70^\circ$ at periapsis), and the red trace has $\chi_0 =
\pi/2$ (so that $\theta = 90^\circ$ at periapsis).  Both traces show
roughly nine complete radial periods.  By their periodic nature, their
motions trace out Lissajous figures: No matter how long we follow
these orbits, they trace out a 1-dimensional trajectory in the
($r,\theta$) plane.

Note that the geometry of the traces in the left-hand panel varies
significantly as $\chi_0$ is varied.  The topology of these traces
remains fixed, however: In all cases the trace oscillates three times
in the angular direction as it completes a single radial oscillation.
As emphasized by Grossman et al.\ {\cite{glpgI}}, the topology of
resonant orbits is uniquely determined by their orbital parameters, by
virtue of the integers $\beta_\theta$ and $\beta_r$ that determine
their periodicity.  We show some evidence in Sec.\ {\ref{sec:results}}
that the topology of resonant orbits directly affects the strength of
their resonance.  Simple orbits, which do not cross themselves often
and do not cover much of the allowed $(r,\theta)$ plane, show large
variations in their radiated fluxes as the phase $\chi_0$ is varied;
more complicated orbits, which cross themselves many times and come
close to much of the allowed $(r,\theta)$ plane, do not show such
large variations.

\section{Gravitational radiation from Kerr orbits}
\label{sec:radiate}

Here we describe in detail how we compute radiation from strong-field
orbits, with an emphasis on how resonances modify the ``usual''
behavior.  We begin in Sec.\ {\ref{sec:fdteuk}} by briefly reviewing
the Teukolsky equation and its solutions.  This material has been
presented at length in several other papers, so we only give a
summary.  Our goal is to provide just enough detail to understand how
the situation changes on resonance.  Section {\ref{sec:fluxes}}
describes how to compute fluxes of energy $E$ and angular momentum
$L_z$ from Teukolsky equation solutions, highlighting how this
calculation must be modified for resonant orbits.  The analogous
calculation for the Carter constant calculation is sufficiently
complicated that we present its details in Appendix {\ref{app:qdot}}.
Finally, Secs.\ {\ref{sec:resI}} and {\ref{sec:resII}} present two
different ways to compute on-resonant fluxes.  These methods are
equivalent to one another, although their computational
implementations are quite different.  As mentioned in the
Introduction, our analytic results for fluxes of $E$ and $L_z$ agree
with those obtained in Ref.\ \cite{glpgII}, while our results for the
Carter constant are new.

\subsection{The frequency-domain Teukolsky equation and its solutions}
\label{sec:fdteuk}

Our computation of the small body's self interaction uses the
Teukolsky equation {\cite{teuk}}.  This equation governs the radiative
components to a Kerr black hole's spacetime curvature, $\psi_0$ and
$\psi_4$, which arise due to some perturbing source or field.  In the
relevant limits, identities make it possible to obtain all information
about the field $\psi_0$ from $\psi_4$, and vice versa, so we need
only focus on one.  The field $\psi_4$ is particularly convenient for
describing radiation at infinity.

Teukolsky showed {\cite{teuk}} that, imposing the Fourier and
multipolar decomposition
\begin{equation}
\psi_4 = \rho^4\int_{-\infty}^{\infty}d\omega \sum_{lm}
R_{lm\omega}(r) S_{lm\omega}(\theta) e^{i(m\phi - \omega t)}\;,
\label{eq:psi4decomp}
\end{equation}
where $\rho = -1/(r - ia\cos\theta)$, a master partial differential
equation governing $\psi_4$ separates.  The function
$S_{lm\omega}(\theta)$ is a spin-weighted spheroidal harmonic;
Ref.\ {\cite{h00}} presents techniques for computing it to high
accuracy.  The radial function is governed by
\begin{equation}
\Delta^2\frac{d}{dr}\left(\frac{1}{\Delta}\frac{dR_{lm\omega}}{dr}\right)
- V(r)R_{lm\omega} = -{\cal T}_{lm\omega}(r,\chi_0)\;.
\label{eq:teuk}
\end{equation}
Equation (\ref{eq:teuk}) is the Teukolsky equation (although that name
is also used for the PDE that governs $\psi_4$ before separating
variables).  Setting the right-hand side of (\ref{eq:teuk}) to zero,
we construct a pair of homogeneous solutions, $R^H_{lm\omega}$ (which
is regular on the event horizon) and $R^\infty_{lm\omega}$ (which is
regular at infinity).  See Ref.\ {\cite{dh06}} (hereafter DH06) for
detailed discussion of how we construct these solutions, as well as
for the potential $V(r)$ appearing in Eq.\ (\ref{eq:teuk}).  From
these solutions, it is straightforward to build a Green's function
which can then be integrated over the source ${\cal T}_{lm\omega}$ to
construct a particular solution.

The source ${\cal T}_{lm\omega}$ is sufficiently complicated that we
will not write it out explicitly; see DH06 for details.  It is
built from projections of the stress-energy tensor for a small body
orbiting the black hole,
\begin{equation}
T_{\alpha\beta} = \frac{\mu\,u_\alpha
  u_\beta}{\Sigma\,\sin\theta\,dt/d\tau} \delta[r-r_{\rm o}(t)]
\delta[\theta-\theta_{\rm o}(t,\chi_0)] \delta[\phi-\phi_{\rm o}(t)]\;,
\end{equation}
where $\mu$ is the mass of the small body, and $u_\alpha$ are the
components of its orbital 4-velocity.  The subscript ``o'' on the
coordinates in the delta functions stands for ``orbit,'' labeling the
orbit's coordinates (as opposed to a general field point, which we
leave without a subscript).

Note that ${\cal T}_{lm\omega}$ is a frequency-domain quantity.
Because it arises from Kerr orbital motion, it only has support at
frequencies $\omega_{mkn} = m\Omega_\phi + k\Omega_\theta +
n\Omega_r$, and is non-zero only for $r_{\rm min} \le r \le r_{\rm
  max}$, $\theta_m \le \theta \le \pi - \theta_m$ [where $r_{\rm rmin}
  = p/(1 + e)$, and $r_{\rm max} = p/(1 - e)$; see
  Eq.\ (\ref{eq:rdefthdef})].  Once fully constructed, ${\cal
  T}_{lm\omega}$ has terms in $\delta[r - r_{\rm o}(t)]$ and its first
two radial derivatives; see Sec.\ III of DH06.

To understand fluxes from this system, our interest is in
$R_{lm\omega}(r)$ in the limits $r \to \infty$ and $r \to r_+$ (the
event horizon).  These limits will allow us to deduce how the orbit
evolves due to radiation to infinity, and due to radiation absorbed by
the hole.  As $r \to \infty$, the homogeneous solution
$R^\infty_{lm\omega}(r)$ approaches (modulo a power-law scaling) an
outgoing plane wave.  Likewise, as $r \to r_+$, the solution
$R^H_{lm\omega}(r)$ limits to an ingoing plane wave.  The particular
solution we construct by integrating the Green's function over the
source then takes the form
\begin{equation}
R_{lm\omega}(r) = \left\{ \begin{array}{ll}
Z^H_{\omega lm}(\chi_0) R^\infty_{\omega lm}(r)
& \mbox{\ \ \ \  $ r \to \infty$,} \\
Z^\infty_{\omega lm}(\chi_0) R^H_{\omega lm}(r)
& \mbox{\ \ \ \ $r \to r_+$}, \\
\end{array}
\right.
\end{equation}
where
\begin{equation}
Z^\star_{lm\omega}(\chi_0) = C^\star \int_{r_+}^\infty
dr'\frac{R^\star_{lm\omega}(r'){\cal
    T}_{lm\omega}(r',\chi_0)}{\Delta(r')^2}\;,
\label{eq:Z1}
\end{equation}
and where $\star$ can stand for $\infty$ or $H$.  The symbol $C^\star$
is shorthand for a collection of constants whose value is not needed
here; see Sec.\ III of DH06 for further discussion.

Next insert ${\cal T}_{lm\omega}$ into Eq.\ (\ref{eq:Z1}) and perform
the $r$ integral.  The result is a Fourier transform:
\begin{eqnarray}
Z^\star_{lm\omega}(\chi_0) &=& C^\star \int_{-\infty}^\infty
dt\,e^{i[\omega t - \phi(t)]} I^\star_{lm\omega}[r_{\rm
    o}(t),\theta_{\rm o}(t,\chi_0)]
\nonumber\\
&=& C^\star \int_{-\infty}^\infty d\lambda\, e^{i(\omega \Gamma - m
  \Upsilon_\phi)\lambda} \times
\nonumber\\
& &\quad\qquad\qquad J^\star_{lm\omega}[r_{\rm o}(\lambda),
  \theta_{\rm o}(\lambda,\chi_0)]\;.
\label{eq:Z2}
\end{eqnarray}
The function $I^\star_{lm\omega}$ introduced on the first line of
Eq.\ (\ref{eq:Z2}) is built from ${\cal T}_{lm\omega}$; see
Eqs.\ (3.30)--(3.33) in DH06 and associated text for detailed
discussion.  On the second line, we have changed the integration
variable from coordinate time $t$ to Mino time $\lambda$, and defined
\begin{eqnarray}
J^\star_{lm\omega}(r_{\rm o},\theta_{\rm o}) &=&
I^\star_{lm\omega}(r_{\rm o},\theta_{\rm o})\,T(r_{\rm o},\theta_{\rm o})
\times
\nonumber\\
& & e^{i[\omega\Delta t(r_{\rm o},\theta_{\rm o}) -
  m\Delta\phi(r_{\rm o},\theta_{\rm o})]}\;.
\label{eq:Jdef}
\end{eqnarray}
[In any place that we indicate a dependence on $(r_{\rm o},\theta_{\rm
    o})$, please note that this is shorthand for $[r_{\rm
      o}(\lambda),\theta_{\rm o}(\lambda,\chi_0)]$.]  The function
$J^\star_{lm\omega}(r_{\rm o},\theta_{\rm o})$ is just
$I_{lm\omega}^\star(r_{\rm o},\theta_{\rm o})$ reweighted by $T(r_{\rm
  o},\theta_{\rm o})$ [the right-hand side of the geodesic equation
  (\ref{eq:tdot})], and with the factor $e^{i(\omega\Delta t -
  m\Delta\phi)}$ included.  The functions $\Delta t(r_{\rm
  o},\theta_{\rm o})$ and $\Delta\phi(r_{\rm o},\theta_{\rm o})$ are
oscillatory contributions to the $t$ and $\phi$ pieces of the orbit:
\begin{eqnarray}
\label{tdecompos}
t_{\rm o}(\lambda) &=& \Gamma \lambda + \Delta t[r_{\rm
    o}(\lambda),\theta_{\rm o}(\lambda,\chi_0)]\;,
\\
\phi_{\rm o}(\lambda) &=& \Upsilon_\phi \lambda + \Delta\phi[r_{\rm
    o}(\lambda), \theta_{\rm o}(\lambda,\chi_0)]\;.
\label{phidecompos}
\end{eqnarray}
Both $\Delta t$ and $\Delta\phi$ oscillate at harmonics of
$\Upsilon_\theta$ and $\Upsilon_r$; see Ref.\ {\cite{dh04}} for
detailed discussion.

The function $J^\star_{lm\omega}(r_{\rm o},\theta_{\rm o})$ gathers
all the pieces of the integrand for $Z^\star_{lm\omega}$ that can be
described as harmonics of $\Upsilon_\theta$ and $\Upsilon_r$.  As
such, it is useful to decompose it into these harmonics:
\begin{equation}
J^\star_{lm\omega}(r_{\rm o},\theta_{\rm o})
= \sum_{kn} J^\star_{\omega lmkn}(\chi_0) e^{-i(k\Upsilon_\theta +
  n\Upsilon_r)\lambda}\;,
\label{eq:Jdecomp}
\end{equation}
where
\begin{eqnarray}
J^\star_{\omega lmkn}(\chi_0) &=&
\frac{\Upsilon_r\Upsilon_\theta}{(2\pi)^2}\int_0^{2\pi/\Upsilon_\theta}
d\lambda^\theta
\int_0^{2\pi/\Upsilon_r}d\lambda^r\,
\nonumber\\
& &
e^{i(k\Upsilon_\theta \lambda^\theta + n\Upsilon_r\lambda^r)}
J^\star_{lm\omega}[r_{\rm o}(\lambda^r),\theta_{\rm
    o}(\lambda^\theta,\chi_0)]\;.
\nonumber\\
\label{eq:Jcoef}
\end{eqnarray}
We have here taken advantage of the fact that Mino time completely
decouples the $r$ and $\theta$ motions from one another.  We imagine
that these two coordinates depend separately on two different
Mino-time variables, $\lambda^r$ and $\lambda^\theta$, and integrate
over a full period of each time.  See Ref.\ {\cite{dh04}} for
detailed discussion of this trick.

Next, combine Eqs.\ (\ref{eq:Z2}), (\ref{eq:Jdef}),
(\ref{eq:Jdecomp}), and (\ref{eq:Jcoef}) to find
\begin{eqnarray}
Z^\star_{lm\omega}(\chi_0) &=&
\frac{2\pi}{\Gamma}\sum_{kn}J^\star_{\omega lmkn}(\chi_0) \delta(\omega -
\omega_{mkn})
\nonumber\\
&\equiv&
\sum_{kn} Z^\star_{lmkn}(\chi_0)\delta(\omega -
\omega_{mkn})\;.
\label{eq:Zexpand}
\end{eqnarray}
On the last line, we have taken advantage of the fact that the delta
functions mean that the RHS only has support at $\omega =
\omega_{mkn}$, and we have defined
\begin{eqnarray}
Z^\star_{\omega lmkn}(\chi_0) &=& \frac{2\pi}{\Gamma}
J^\star_{\omega lmkn}(\chi_0)
\label{eq:Zcoef1}\\
&=& \frac{\Upsilon_r\Upsilon_\theta}{2\pi\Gamma}
\int_0^{2\pi/\Upsilon_\theta}d\lambda^\theta
\int_0^{2\pi/\Upsilon_r}d\lambda^r \nonumber\\
& &\!\!\!\!\!  e^{i(k\Upsilon_\theta\lambda^\theta +
  n\Upsilon_r\lambda^r)} J^\star_{lm\omega}[r_{\rm
    o}(\lambda^r),\theta_{\rm o}(\lambda^\theta,\chi_0)]
\nonumber\\
\label{eq:Zcoef}
\end{eqnarray}
and
\begin{eqnarray}
Z^\star_{lmkn}(\chi_0) = Z^\star_{\omega_{mkn}lmkn}(\chi_0).
\label{eq:Zcoefa}
\end{eqnarray}

Throughout this synopsis, we have explicitly shown the dependence on
the relative phase $\chi_0$.  To account for its influence on the
amplitudes, let us first define
\begin{equation}
\check{Z}^\star_{lmkn} \equiv Z^\star_{lmkn}(\chi_0=0)\;.
\end{equation}
In other words, amplitudes with a check mark $\check{\ }$ are computed
using the fiducial geodesic.  As shown in Sec.\ 8.4 of
Ref.\ {\cite{dfh05}}, the effect of $\chi_0$ is to introduce a phase:
\begin{equation}
Z^\star_{lmkn}(\chi_0) = e^{i\xi_{mkn}(\chi_0)} \check{Z}^\star_{lmkn}\;,
\label{eq:Zphaseshift}
\end{equation}
where
\begin{eqnarray}
\xi_{mkn}(\chi_0) &=& k\Upsilon_\theta \lambda^\theta_0 +
m\Delta\hat\phi[r_{\rm min}, \theta(-\lambda^\theta_0)]
\nonumber\\
& &
\quad - \omega_{mkn}\Delta\hat t[r_{\rm min},
  \theta(-\lambda^\theta_0)]\;,
\label{eq:xi_val}
\end{eqnarray}
where $\Delta\hat\phi$ is $\Delta\phi$ for the fiducial geodesic (and
likewise for $\Delta\hat t$), and where $\lambda^\theta_0 =
\lambda^\theta_0(\chi_0)$ is the value of $\lambda^\theta$ at which
$\theta = \theta_m$.  It is given explicitly by Eq.\ (3.75) of
Ref.\ \cite{dfh05}.  On the fiducial geodesic, $\lambda^\theta_0 = 0$,
and $\xi_{mkn} = 0$, as it should.

\subsection{The non-resonant rates of change of the orbital
parameters $E$, $L_z$, and $Q$}
\label{sec:fluxes}

As stated previously, our eventual goal is to compute the motion of a
body which spirals through resonances under a rigorously computed self
force, or at least the dissipative piece of the self force.  The three
components of the self force can be regarded as the rates of change of
the orbital constants $E$, $L_z$, $Q$.  We will present results
showing these rates of change for the dissipative self force in a
later paper.  Here, we focus just on appropriately averaged rates of
change of $E$, $L_z$, and $Q$.

In this section, we will how to extract the rates at which
gravitational radiation carries $E$ and $L_z$ to infinity and down the
event horizon.  This calculation has appeared many times in other
papers; we present it in perhaps more detail than is necessary in
order to highlight aspects of the calculation that change when we move
from non-resonant to resonant orbits.  One cannot extract the rate of
change of $Q$ from the radiation, but must instead compute it using
the dissipative self force.  This is was done by Sago et
al.\ {\cite{sago}} (hereafter S06).  We go through the Sago et
al.\ calculation in some detail in Appendix {\ref{app:qdot}} in order
to understand how to modify their result on an orbital resonance.  In
Appendix {\ref{app:edotlzdot}}, we likewise compute the rates of
change of $E$ and $L_z$ using the dissipative self force.  The result
we find there (for both resonant and non-resonant orbits) duplicates
the rates of change we compute from gravitational-wave fluxes.  This
is not terribly surprising: Quinn and Wald {\cite{qw99}} showed that
this equality must hold given an appropriate averaging for these two
ways of computing the evolution of $E$ and $L_z$.  Strictly speaking,
Quinn and Wald's does not apply to the situation we are studying ---
they do not consider black hole spacetimes (although they describe how
to go beyond their calculation to include this limit), and require
that the particle's trajectory begin and end far away from the
gravitating source.  Nonetheless, it demonstrates that this balance is
to be expected in a wide range of situations, so the equality we find
is sensible.

Using Eq.\ (\ref{eq:psi4decomp}) and the definitions which follow, we
find that as $r \to \infty$,
\begin{eqnarray}
\psi_4 &=& \frac{1}{r}\sum_{lmkn} e^{i\xi_{mkn}(\chi_0)} \check{Z}^H_{lmkn}
S_{lmkn}(\theta)
e^{i(m\phi - \omega_{mkn}t)}\nonumber\\
&\equiv& \frac{1}{r}\sum_{lmkn} \psi_{4,lmkn}\;.
\label{eq:psi4decomp2}
\end{eqnarray}
Here, $S_{lmkn}(\theta)$ is the spheroidal harmonic
$S_{lm\omega}(\theta)$ for $\omega = \omega_{mkn}$.  As $r \to
\infty$, $\psi_4 \to (1/2)(\ddot h_+ - i\ddot h_\times)$, so
\begin{equation}
h_+ - ih_\times =
-\frac{2}{r}\sum_{lmkn}\frac{\psi_{4,lmkn}}{\omega_{mkn}^2}\;.
\label{eq:psi4asymptotic}
\end{equation}

A useful tool for understanding the energy carried by gravitational
waves is the Isaacson stress-energy tensor {\cite{isaacson}}, whose $r
\to \infty$ limit is given by
\begin{eqnarray}
T^{\rm rad}_{\mu\beta}
%&=& \frac{1}{16\pi} \left\langle \nabla_\mu
%h_+\nabla_\beta h_+ + \nabla_\mu h_\times\nabla_\beta h_\times
%\right\rangle
%\nonumber\\
&=& \frac{1}{16\pi} \left\langle \partial_\mu h_+\partial_\beta h_+
+ \partial_\mu h_\times\partial_\beta h_\times \right\rangle\;.
\label{eq:isaacson}
\end{eqnarray}
The angle brackets in this expression mean that the quantity is
averaged over several wavelengths.  See Ref.\ {\cite{isaacson}} and
references therein for detailed discussion of the averaging procedure.
%  On the first line, $\nabla_\mu$ denotes a covariant derivative
%with respect to the background; on the second line, we take advantage
%of the fact that this limits to the partial derivative as $r \to
%\infty$.

The energy flux, our focus here, is given by
\begin{eqnarray}
\frac{dE^\infty}{dt} &=& \lim_{r\to\infty}r^2\int T^{\rm rad}_{tk}n^k d\Omega
\nonumber\\
&=& \lim_{r\to\infty}r^2\int T^{\rm rad}_{tt} d\Omega\;,
\label{eq:energyflux}
\end{eqnarray}
where $n^k$ is a radially outward pointing normal vector, and the
index $k$ is restricted to spatial directions.

Combining Eqs.\ (\ref{eq:psi4asymptotic}) -- (\ref{eq:energyflux}),
we find
\begin{equation}
\left\langle\frac{dE^\infty}{dt}\right\rangle = \left\langle
\sum_{lmkn}\sum_{l'm'k'n'}{\rm Re} \int \frac{\psi_{4,lmkn}
  \bar\psi_{4,l'm'k'n'}}{4\pi\omega_{mkn}\omega_{m'k'n'}}d\Omega
\right\rangle\;;
\end{equation}
$\bar\psi_4$ is the complex conjugate of $\psi_4$.  The sum over $l$
is taken from $2$ to $\infty$; the sum over $m$ from $-l$ to $l$; the
sums over $k$ and $n$ are both taken from $-\infty$ to $\infty$; and
likewise for the primed indices.  The angle brackets on the left-hand
side mean that this rate of change is to be understood as one which is
averaged over appropriate orbital timescales.

Consider now averaging the right-hand side over several wavelengths.
{\it Assuming that each frequency $\omega_{mkn}$ is distinct} (an
assumption that is only true when we are not on a resonance), then
this averaging forces $m = m'$, $k = k'$, $n = n'$.  Using the fact
that
\begin{equation}
\int S_{lmkn}(\theta)S_{l'mkn}(\theta)d\Omega = \delta_{ll'}\;,
\end{equation}
we find
\begin{equation}
\left\langle\frac{dE^\infty}{dt}\right\rangle = \sum_{lmkn}
\frac{|\check{Z}^H_{lmkn}|^2}{4\pi\omega_{mkn}^2}
\equiv \sum_{lmkn} \dot E^\infty_{lmkn}\;.
\label{eq:EdotInfnonres}
\end{equation}
A similar calculation focusing on $T^{\rm rad}_{t\phi}$ gives us the
flux of axial angular momentum:
\begin{equation}
\left\langle\frac{dL^\infty_z}{dt}\right\rangle = \sum_{lmkn}
\frac{m|\check{Z}^H_{lmkn}|^2}{4\pi\omega_{mkn}^3} \equiv \sum_{lmkn}
\dot L^\infty_{z,lmkn}\;.
\label{eq:LzdotInfnonres}
\end{equation}
Notice that the phase $\xi_{mkn}$ does not appear in
Eqs.\ (\ref{eq:EdotInfnonres}) and (\ref{eq:LzdotInfnonres}).
Appendix {\ref{app:edotlzdot}} derives these results using the local
self force, following S06.

The calculation of fluxes down the horizon is more complicated.  Since
the Isaacson tensor is not defined in a black hole's strong field, we
use the fact that the curvature perturbation from the orbiting body
exerts a shear on the generators of the horizon, which increases the
black hole's surface area.  By the first law of black hole dynamics,
this in turn changes its mass and angular momentum; see
Refs.\ {\cite{teukpress,hh}} for detailed discussion.  Assuming flux
balance, we can then read out the down-horizon fluxes:
\begin{eqnarray}
\left\langle\frac{dE^H}{dt}\right\rangle &=& \sum_{lmkn} \alpha_{lmkn}
\frac{|\check{Z}^\infty_{lmkn}|^2}{4\pi\omega_{mkn}^2}
\equiv \sum_{lmkn}\dot E^H_{lmkn}
\;,
\nonumber\\
\label{eq:EdotHnonres}\\
\left\langle\frac{dL^H_z}{dt}\right\rangle &=& \sum_{lmkn}
\alpha_{lmkn} \frac{m|\check{Z}^\infty_{lmkn}|^2}{4\pi\omega_{mkn}^3}
\equiv \sum_{lmkn}\dot L^H_{z,lmkn}\;.
\nonumber\\
\label{eq:LzdotHnonres}
\end{eqnarray}
We refer the reader to Eq.\ (3.60) of DH06 for the down-horizon factor
$\alpha_{lmkn}$.

Unlike the energy and axial angular momentum, there is no simple
formula describing the ``flux'' of Carter constant carried by
radiation.  However, one can formulate how $Q$ changes due to
radiative backreaction.  Taking into account only the dissipative
piece of the self force and averaging over very long times, Sago et
al. {\cite{sago}} (hereafter S06) showed that
\begin{eqnarray}
\left\langle\frac{dQ^\infty}{dt}\right\rangle &=&
\sum_{lmkn}|\check{Z}^H_{lmkn}|^2 \times \frac{\left({\cal L}_{mkn} +
  k\Upsilon_\theta\right)}{2\pi\omega_{mkn}^3} \;,
\label{eq:qdotInf}\\
\left\langle\frac{dQ^H}{dt}\right\rangle &=&
\sum_{lmkn}\alpha_{lmkn}|\check{Z}^\infty_{lmkn}|^2\times\frac{\left({\cal
    L}_{mkn} + k\Upsilon_\theta\right)}{2\pi\omega_{mkn}^3}\;,
\nonumber\\
\label{eq:qdotH}
\end{eqnarray}
where
\begin{equation}
{\cal L}_{mkn} = m\langle\cot^2\theta\rangle L_z - a^2\omega_{mkn}
\langle\cos^2\theta\rangle E\;.
\label{eq:calLdef}
\end{equation}
It is interesting that the rate of change of $Q$ can be factored into
quantities that are encoded in the distant radiation
($\check{Z}^H_{lmkn}$ and $\check{Z}^\infty_{lmkn}$) and quantities
that are local to the orbital worldline (${\cal L}_{mkn}$,
$\omega_{mkn}$, and $\Upsilon_\theta$).  Using
Eqs.\ (\ref{eq:EdotInfnonres}) and (\ref{eq:EdotHnonres}), these
results can be written
\begin{eqnarray}
\left\langle\frac{dQ^\star}{dt}\right\rangle = 2\sum_{lmkn} \dot
E^\star_{lmkn} \times \left({\cal L}_{mkn} + k\Upsilon_\theta\right)
/\omega_{mkn}\;,\nonumber\\
\end{eqnarray}
where $\star$ is either $\infty$ or $H$.  We go through the Sago et
al.\ calculation of $\langle dQ/dt\rangle$ in some detail in Appendix
{\ref{app:qdot}} in order to understand how to modify this result on
an orbital resonance.

Note that the rates of change $\langle dE^\star/dt\rangle$, $\langle
dL_z^\star/dt\rangle$, and $\langle dQ^\star/dt\rangle$ are equivalent
for non-resonant orbits to the three components of the torus-averaged
forcing term $\langle G^{(1)}_i\rangle$ introduced in the
introduction, albeit using coordinate time $t$ rather than proper time
$\tau$ to parameterize the rate of change.  This equivalence breaks
down for resonant orbits, as pointed out in Ref.\ \cite{glpgII}.

\subsection{Radiation from resonant orbits I: Merging of amplitudes
on resonance}
\label{sec:resI}

On resonance, $\Omega_\theta/\beta_\theta = \Omega_r/\beta_r \equiv
\Omega$, and so $k\Omega_\theta + n\Omega_r = N\Omega$, where $N =
k\beta_\theta + n\beta_r$.  An infinite number of pairs $(k,n)$ are
consistent with a given $N$.  For a given value of $m$, all pairs
$(k,n)$ satisfying $k\beta_\theta + n\beta_r = N$ will have mode
frequency $\omega_{mkn} \equiv \omega_{mN} = m\Omega_\phi + N\Omega$.

Revisiting Eq.\ (\ref{eq:psi4decomp2}), this means that only three
indices are needed to describe the radiation on resonance, rather
than four:
\begin{equation}
\psi_4^{\rm res} = \frac{1}{r}\sum_{lmN}{\cal Z}_{lmN}^H(\chi_0)
S_{lmN}(\theta)e^{i(m\phi - \omega_{mN}t)}\;,
\label{eq:psi4decomp_res}
\end{equation}
where
\begin{equation}
{\cal Z}^\star_{lmN}(\chi_0) = \sum_{(k,n)_N} e^{i\xi_{mkn}(\chi_0)}
\check{Z}^\star_{lmkn}\;,
\label{eq:Zres_expand}
\end{equation}
and where $(k,n)_N$ denotes all pairs $(k,n)$ which satisfy
$k\beta_\theta + n\beta_r = N$.  In Eq.\ (\ref{eq:psi4decomp_res}),
the sums over $l$ and $m$ are exactly as before, and $N$ is summed
from $-\infty$ to $\infty$.  Equation (C16) of Ref.\ {\cite{glpgII}}
gives a relationship, in their notation, that is equivalent to our
Eq.\ (\ref{eq:Zres_expand}).

Equations (\ref{eq:psi4decomp_res}) and (\ref{eq:Zres_expand}) tell us
that, as we enter a resonance, modes of $\psi_4$ which were distinct
combine with one another: ``lines'' in the gravitational-wave spectrum
merge.  Each mode's contribution to the combined amplitude
(\ref{eq:Zres_expand}) is weighted by its phase $\xi_{mkn}(\chi_0)$.
Revisiting the calculation of the fluxes using
Eq.\ (\ref{eq:psi4decomp_res}) rather than (\ref{eq:psi4decomp2}), we
find
\begin{eqnarray}
\left\langle\frac{dE^\infty}{dt}(\chi_0)\right\rangle &=& \sum_{lmN}
\frac{|{\cal Z}^H_{lmN}(\chi_0)|^2}{4\pi\omega_{mN}^2}
\nonumber\\
&\equiv& \sum_{lmN} \dot E^\infty_{lmN}(\chi_0)\;,
\label{eq:edotInf_res}\\
\left\langle\frac{dE^H}{dt}(\chi_0)\right\rangle &=& \sum_{lmN}
\alpha_{lmN} \frac{|{\cal
    Z}^\infty_{lmN}(\chi_0)|^2}{4\pi\omega_{mN}^2}
\nonumber\\
&\equiv& \sum_{lmN} \dot E^H_{lmN}(\chi_0)\;,
\label{eq:edotH_res}\\
\left\langle\frac{dL^\infty_z}{dt}(\chi_0)\right\rangle
&=& \sum_{lmN}
\frac{m|{\cal Z}^\infty_{lmN}(\chi_0)|^2}{4\pi\omega_{mN}^3}
\nonumber\\
&\equiv& \sum_{lmN} \dot L_{z,lmN}^\infty(\chi_0)\;,
\label{eq:lzdotInf_res}\\
\left\langle\frac{dL^H_z}{dt}(\chi_0)\right\rangle &=&
\sum_{lmN}\alpha_{lmN} \frac{m|{\cal
    Z}^\infty_{lmN}(\chi_0)|^2}{4\pi\omega_{mN}^3}
\nonumber\\
&\equiv& \sum_{lmN} \dot L_{z,lmN}^H(\chi_0)\;.
\label{eq:lzdotH_res}
\end{eqnarray}
(The factor $\alpha_{lmN}$ appearing here is the same as
$\alpha_{lmkn}$ introduced earlier, but with $\omega_{mkn}$ replaced
by $\omega_{mN}$.)  Thanks to the dependence of ${\cal Z}^\star_{lmN}$
on the relative phase $\chi_0$, the on-resonance fluxes likewise
depend on this phase.  These equations reproduce Eq.\ (C15) of
Ref.\ {\cite{glpgII}}.  We derive them using the local self force in
Appendix {\ref{app:edotlzdot}}.

In Appendix {\ref{app:qdot}}, we show how the calculation of $dQ/dt$
is changed due to an orbital resonance.  The result is
\begin{eqnarray}
\left\langle\frac{dQ^\infty}{dt}(\chi_0)\right\rangle = \sum_{lmN}
\frac{|{\cal Z}^H_{lmN}(\chi_0)|^2}{2\pi\omega_{mN}^3}{\cal L}_{mN}
\qquad\qquad
\nonumber\\
\quad + \Upsilon_\theta\sum_{lmN}\frac{\mbox{Re}\left[{\cal
      Z}^H_{lmN}(\chi_0)\bar{\cal
      Y}^H_{lmN}(\chi_0)\right]}{2\pi\omega_{mN}^3}\;,
\nonumber\\
\label{eq:Qdotres_Inf}\\
\left\langle\frac{dQ^H}{dt}(\chi_0)\right\rangle = \sum_{lmN}
\frac{\alpha_{lmN}|{\cal
    Z}^\infty_{lmN}(\chi_0)|^2}{2\pi\omega_{mN}^3}{\cal L}_{mN}
\qquad\qquad \nonumber\\
\quad + \Upsilon_\theta\sum_{lmN}\frac{\alpha_{lmN}\mbox{Re}
  \left[{\cal Z}^\infty_{lmN}(\chi_0)\bar{\cal
      Y}^\infty_{lmN}(\chi_0)\right]}{2\pi\omega_{mN}^3}\;.
\nonumber\\
\label{eq:Qdotres_H}
\end{eqnarray}
The factor ${\cal L}_{mN}$ is the same as ${\cal L}_{mkn}$ with
$\omega_{mkn}$ replaced by $\omega_{mN}$.  We have introduced the
modified amplitude
\begin{equation}
{\cal Y}^\star_{lmN}(\chi_0) = \sum_{(k,n)_N} k e^{i\xi_{mkn}(\chi_0)}
\check{Z}^\star_{lmkn}\;.
\label{eq:Ycof_def}
\end{equation}
Notice that ${\cal Y}^\star_{lmN}(\chi_0)$ is similar to ${\cal
  Z}^\star_{lmN}(\chi_0)$ [compare Eq.\ (\ref{eq:Zres_expand})], but
with each term in the sum weighted by $k$.  Equations
(\ref{eq:Qdotres_Inf}) and (\ref{eq:Qdotres_H}) are used in the
following section to study how the Carter constant's evolution is
affected by an orbital resonance.

\subsection{Radiation from resonant orbits II: The constrained
source integral of a resonant orbit}
\label{sec:resII}

The method described in Sec.\ {\ref{sec:resI}} builds the on-resonance
amplitudes ${\cal Z}^\star_{lmN}(\chi_0)$ from the amplitudes
$\check{Z}^\star_{lmkn}$ which are normally computed with
frequency-domain Teukolsky equation solvers, such as that described in
DH06.  The only modification is the need to compute the phase
$\xi_{mkn}(\chi_0)$.

One can also compute the on-resonant amplitudes by modifying the
integral for the amplitudes $\check{Z}^\star_{lmkn}$.  Doing so, we
compute ${\cal Z}^\star_{lmN}(\chi_0)$ directly, without reference to
the amplitudes $\check{Z}^\star_{lmkn}$.  We begin this calculation by
carrying over without modification the computation of
Sec.\ {\ref{sec:fdteuk}} up to Eq.\ (\ref{eq:Z2}),
\begin{equation*}
Z^\star_{lm\omega} = C^\star \int_{-\infty}^\infty d\lambda\,
e^{i(\omega \Gamma - m \Upsilon_\phi)\lambda}
J^\star_{lm\omega}[r_{\rm o}(\lambda), \theta_{\rm o}(\lambda,\chi_0)]\;.
\end{equation*}
As before, we decompose $J^\star_{lm\omega}$ into $\Upsilon_\theta$
and $\Upsilon_r$ harmonics.  However, we now take into account how
these frequencies are related on a resonance:
\begin{eqnarray}
J^\star_{lm\omega} &=& \sum_{kn}J^\star_{\omega lmkn}
e^{-i(k\Upsilon_\theta + n\Upsilon_r)\lambda}
\nonumber\\
&=& \sum_{kn}J^\star_{\omega lmkn} e^{-i(k\beta_\theta + n\beta_r)\Upsilon\lambda}
\nonumber\\
&\equiv& \sum_{N}{\cal J}^\star_{\omega lmN} e^{-iN\Upsilon\lambda}\;.
\label{eq:Jdecomp_res}
\end{eqnarray}
On the second line, we've used the resonance relation
$\Upsilon_\theta/\beta_\theta = \Upsilon_r/\beta_r \equiv \Upsilon$.
We then use $N = k\beta_\theta + n\beta_r$, and change notation
slightly to distinguish the source amplitude $J^\star_{\omega lmkn}$
from its on-resonance variant ${\cal J}^\star_{\omega lmN}$.

The result, Eq.\ (\ref{eq:Jdecomp_res}), depends on only one
fundamental frequency, $\Upsilon$.  As such, our integral for ${\cal
  J}^\star_{\omega lmN}$ is taken over only a single time variable
$\lambda$:
\begin{equation}
{\cal J}^\star_{\omega lmN}(\chi_0) = \frac{\Upsilon}{2\pi}
\int_0^{2\pi/\Upsilon} d\lambda\, J^\star_{lm\omega}[r_{\rm
    o}(\lambda),\theta_{\rm o}(\lambda,\chi_0)]
e^{iN\Upsilon\lambda}\;.
\label{eq:Jcoef_res}
\end{equation}
Finally, by combining Eqs.\ (\ref{eq:Z2}), (\ref{eq:Jdef}),
(\ref{eq:Jdecomp_res}), and (\ref{eq:Jcoef_res}), we define
\begin{eqnarray}
{\cal Z}^\star_{lmN}(\chi_0) &=&
\frac{2\pi}{\Gamma}{\cal J}^\star_{\omega_{mN}lmN}(\chi_0)
\label{eq:Zres_int1}\\
&=& \frac{\Upsilon}{\Gamma}\int_0^{2\pi/\Upsilon}d\lambda\,
J^\star_{lm\omega_{mN}}[r(\lambda),\theta(\lambda,\chi_0)]e^{i N
  \Upsilon\lambda}\;.
\nonumber\\
\label{eq:Zres_int}
\end{eqnarray}
Equation (B33) of Ref.\ {\cite{glpgII}} is equivalent to
Eq.\ (\ref{eq:Zres_int}) here.

In combining the previous relations to derive
Eq.\ (\ref{eq:Zres_int}), we find a proportionality to $\delta(\omega
- \omega_{mN})$, which forces the RHS to have support only at $\omega
= \omega_{mN}$.  Although it may not be obvious,
Eqs.\ (\ref{eq:Zres_expand}) and (\ref{eq:Zres_int}) are equivalent.
We show this analytically in Appendix {\ref{app:equivalence}}, and
will demonstrate it numerically in the following section.  A
conceptually attractive feature of Eq.\ (\ref{eq:Zres_int}) is that
the integrand is only evaluated at the coordinates $(r,\theta)$ which
the on-resonance orbit passes through.  Changing $\chi_0$ changes the
points $(r,\theta)$ at which the integrand has support.  This is how
the dependence on $\chi_0$ enters ${\cal Z}^\star_{lmN}$ in this
calculation.

However, Eq.\ (\ref{eq:Zres_int}) can only be used for orbits that are
{\it exactly} on resonance.  Indeed, in any other case, the 3-index
amplitude ${\cal Z}^\star_{lmN}$ is not meaningful since the
on-resonance condition $k\beta_\theta + n\beta_r = N$ is not met.  A
suitable generalization of Eq.\ (\ref{eq:Zres_expand}) for slightly
off-resonance orbits can be used to understand the behavior of
$\psi_4$ as one approaches and moves through a resonance.  As such,
the sum of phase-weighted amplitudes, Eq.\ (\ref{eq:Zres_expand}), is
likely to be more useful for understanding the resonant self
interaction in full inspiral studies.  In any case, we have found
having two techniques for computing ${\cal Z}^\star_{lmN}(\chi_0)$ to
be very useful.  The codes which implement these two formulae are
quite different, so it is reassuring that their results are in
agreement.  As discussed at the end of Appendix \ref{app:qdot}, it
appears that the modified amplitude ${\cal Y}_{lmN}(\chi_0)$ can also
be computed with a one dimensional integral by propagating the
operator $(d\theta/d\lambda)\partial_\theta$ under the integral in
Eq.\ (\ref{eq:Zres_int}).  We have not yet tested this, though it
would be a worthwhile exercise to do so.

\section{Results: How resonances impact radiation}
\label{sec:results}

\subsection{Variation of modes with $\chi_0$, and comparison of two
computational techniques}
\label{sec:modeflux}

We now discuss examples illustrating how wave amplitudes and fluxes
are affected by orbital resonances.  All of our results are computed
using a version of the code described in DH06, modified to handle
resonances.

\begin{figure*}[ht]
\includegraphics[width = 0.48\textwidth]{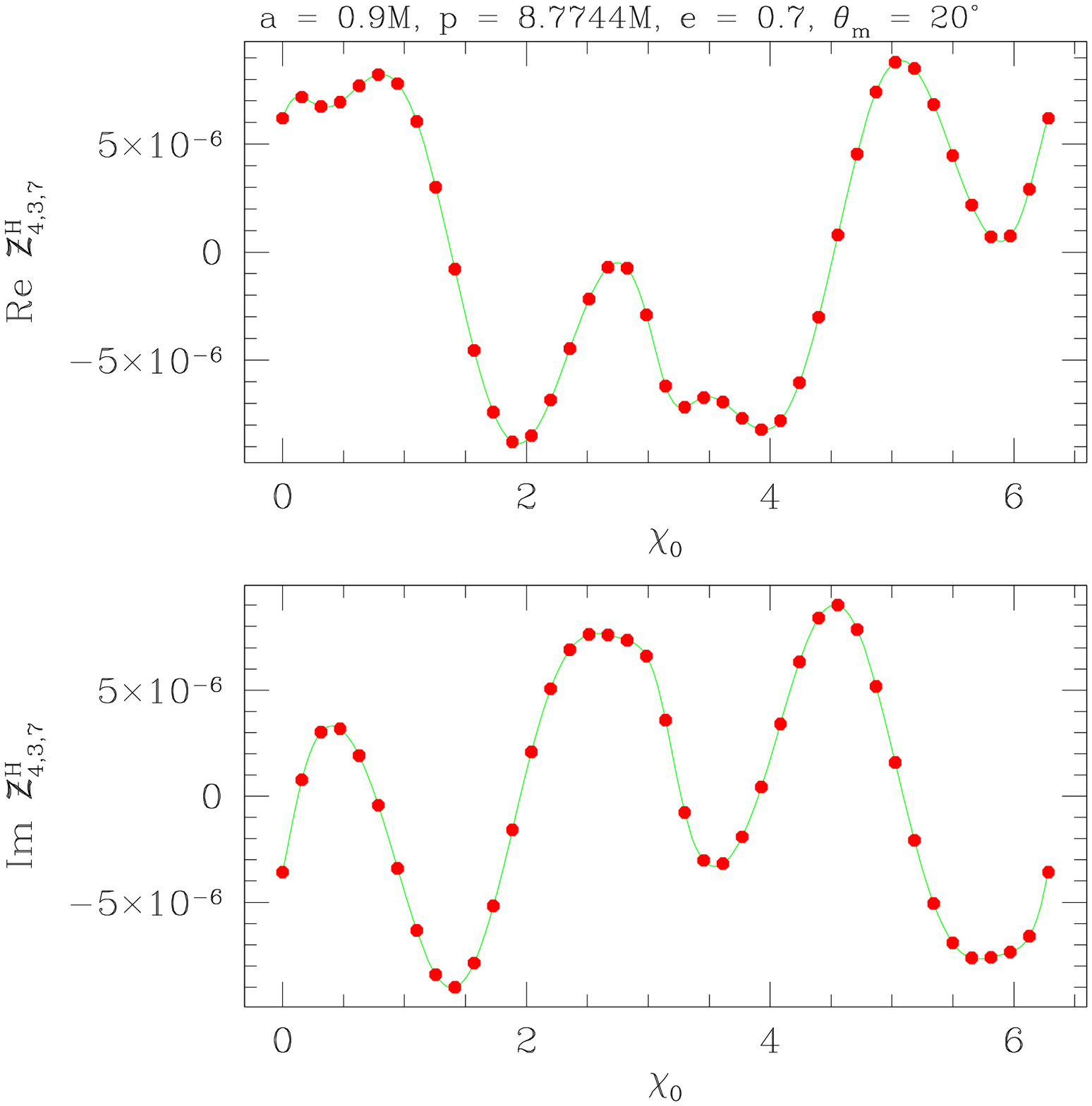}
\includegraphics[width = 0.48\textwidth]{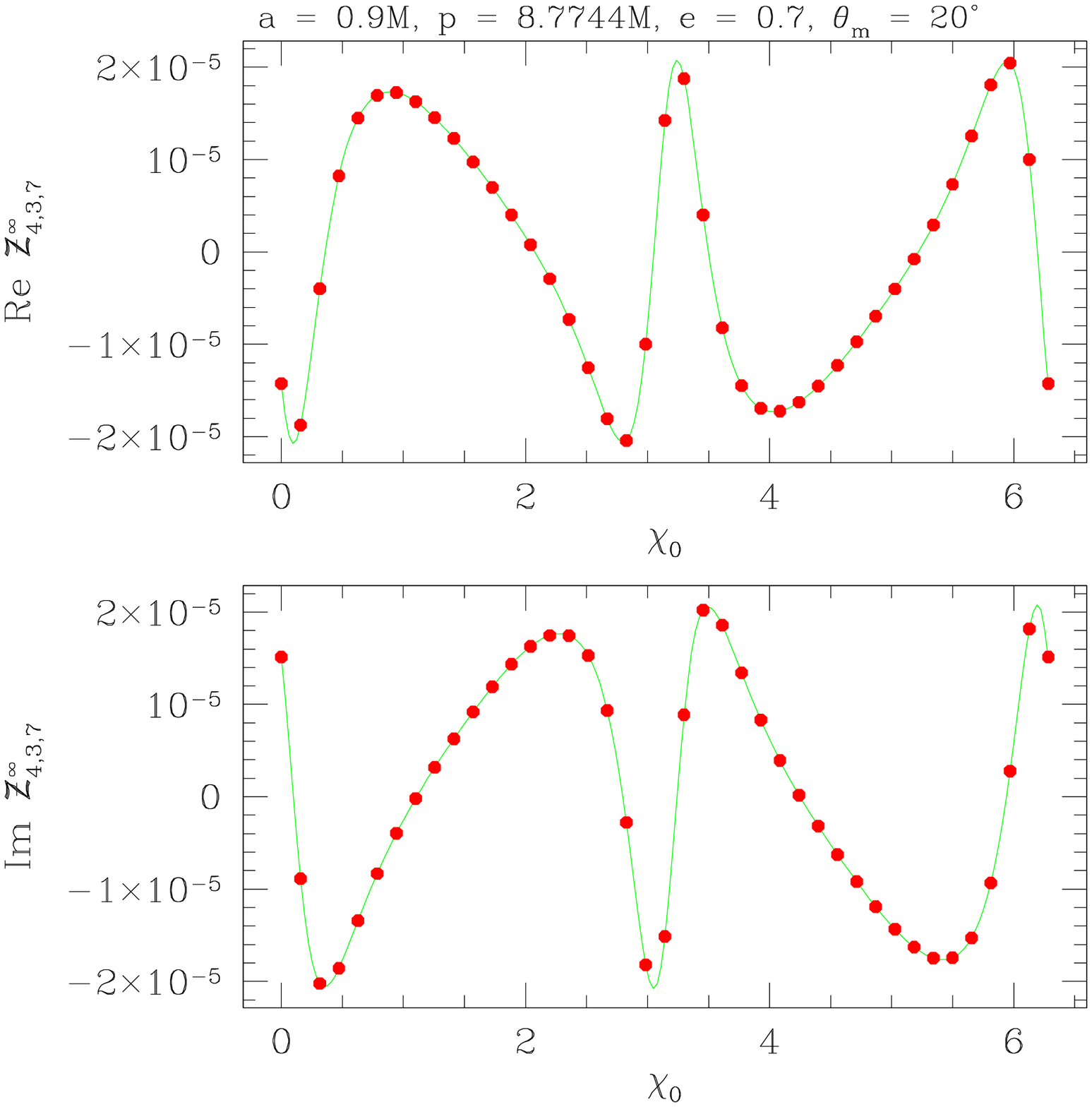}
\caption{Comparison of two methods to compute the on-resonance
  amplitudes ${\cal Z}^\star_{lmN}$.  All panels correspond to
  radiation from an orbit with parameters $p = 8.7744M$, $e = 0.7$,
  $\theta_m = 20^\circ$, about a black hole with spin $a = 0.9M$.  For
  this orbit, $\Omega_\theta/\Omega_r = 3/2$.  We have chosen $l = 4$,
  $m = 3$, $N = 7$.  Left panels show ${\cal Z}^H_{437}$, right panels
  show ${\cal Z}^\infty_{437}$; top panels show the real part, bottom
  panels the imaginary part.  Blue curves show the amplitude computed
  by the on-resonance merging of amplitudes discussed in
  Sec.\ {\ref{sec:resI}}; red dots show the amplitude computed using
  the constrained source integral presented in
  Sec.\ {\ref{sec:resII}}.  The two methods agree to numerical
  accuracy (roughly 6 digits in this case).}
\label{fig:Zeds_compare}
\end{figure*}

Begin with Fig.\ {\ref{fig:Zeds_compare}}, which illustrates how
${\cal Z}^H_{lmN}$ and ${\cal Z}^\infty_{lmN}$ behave as functions of
$\chi_0$.  For this example, we have put $a = 0.9M$, $p = 8.7744M$, $e
= 0.7$, $\theta_m = 20^\circ$ (for which $\Omega_\theta/\Omega_r =
3/2$), and we have chosen $l = 4$, $m = 3$, $N = 7$.  In all panels,
the green curves show ${\cal Z}^\star_{lmN}$ computed using
Eq.\ (\ref{eq:Zres_expand}); the red dots show the same quantity
computed using Eq.\ (\ref{eq:Zres_int}).  The two methods agree to
numerical accuracy (roughly 6 digits\footnote{It is not difficult to
  do the calculations more accurately than this
  {\cite{fujtag1,fujtag2,throweetal}}, but 6 digits of accuracy is
  good enough for this first strong-field examination of this
  effect.}).  All examples that we have examined show that
Eqs.\ (\ref{eq:Zres_expand}) and (\ref{eq:Zres_int}) agree perfectly
(as we would expect from the calculation presented in Appendix
{\ref{app:equivalence}}).  Having both methods at hand was quite
useful for debugging the on-resonance version of our code.

Besides showing the excellent agreement between our methods of
computing ${\cal Z}^\star_{lmN}$, Fig.\ {\ref{fig:Zeds_compare}} also
illustrates how ${\cal Z}^\star_{lmN}$ varies with $\chi_0$.  For this
example, we find that $|{\cal Z}^H_{lmN}|$ varies by about $25\%$ from
minimum to maximum, and $|{\cal Z}^\infty_{lmN}|$ varies by about
$40\%$.  The associated energy fluxes, which are proportional to the
amplitude's modulus squared, varies by about $55\%$ and by a factor of
two, respectively.

\begin{figure*}[ht]
\includegraphics[width = 0.48\textwidth]{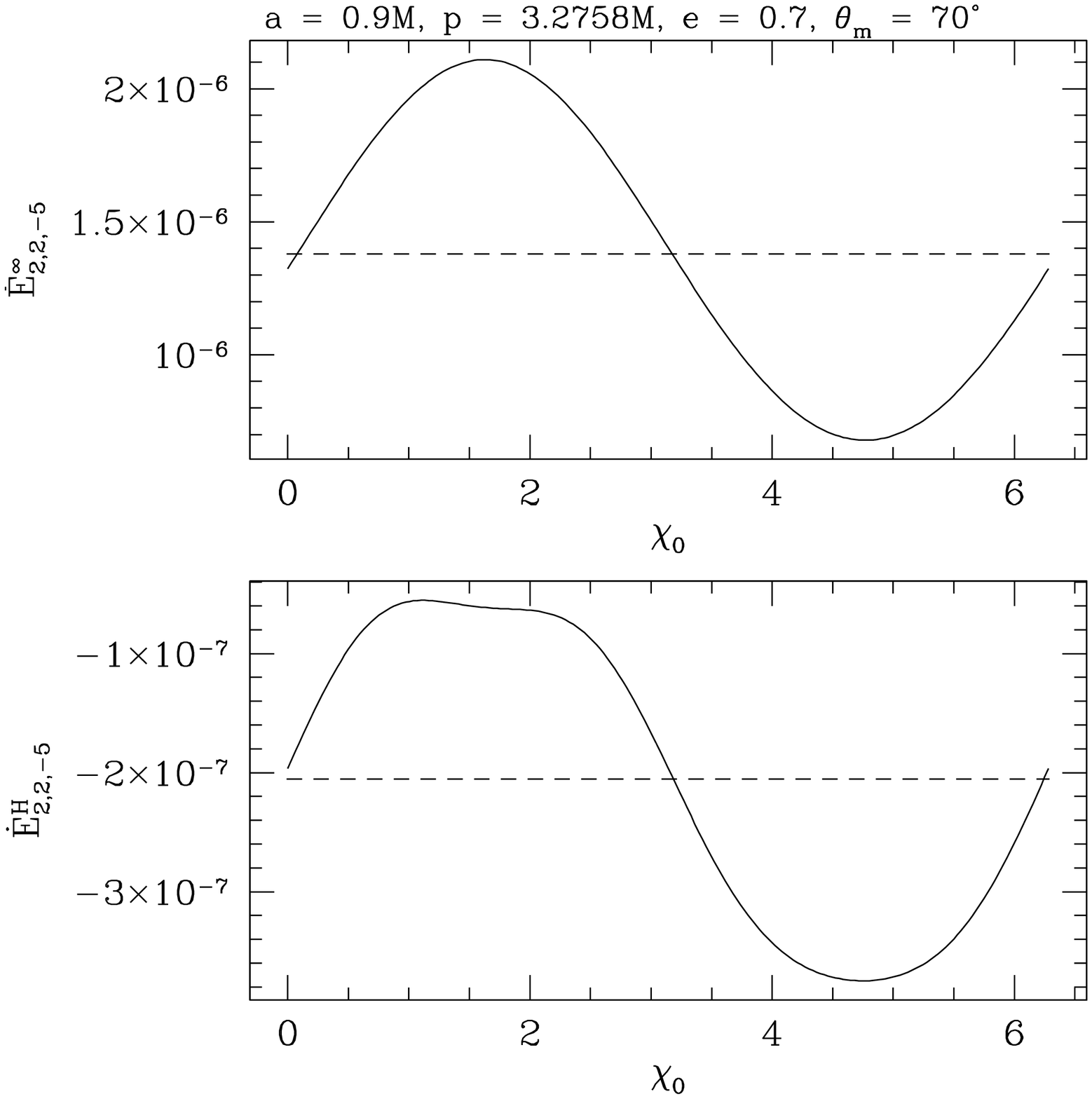}
\includegraphics[width = 0.48\textwidth]{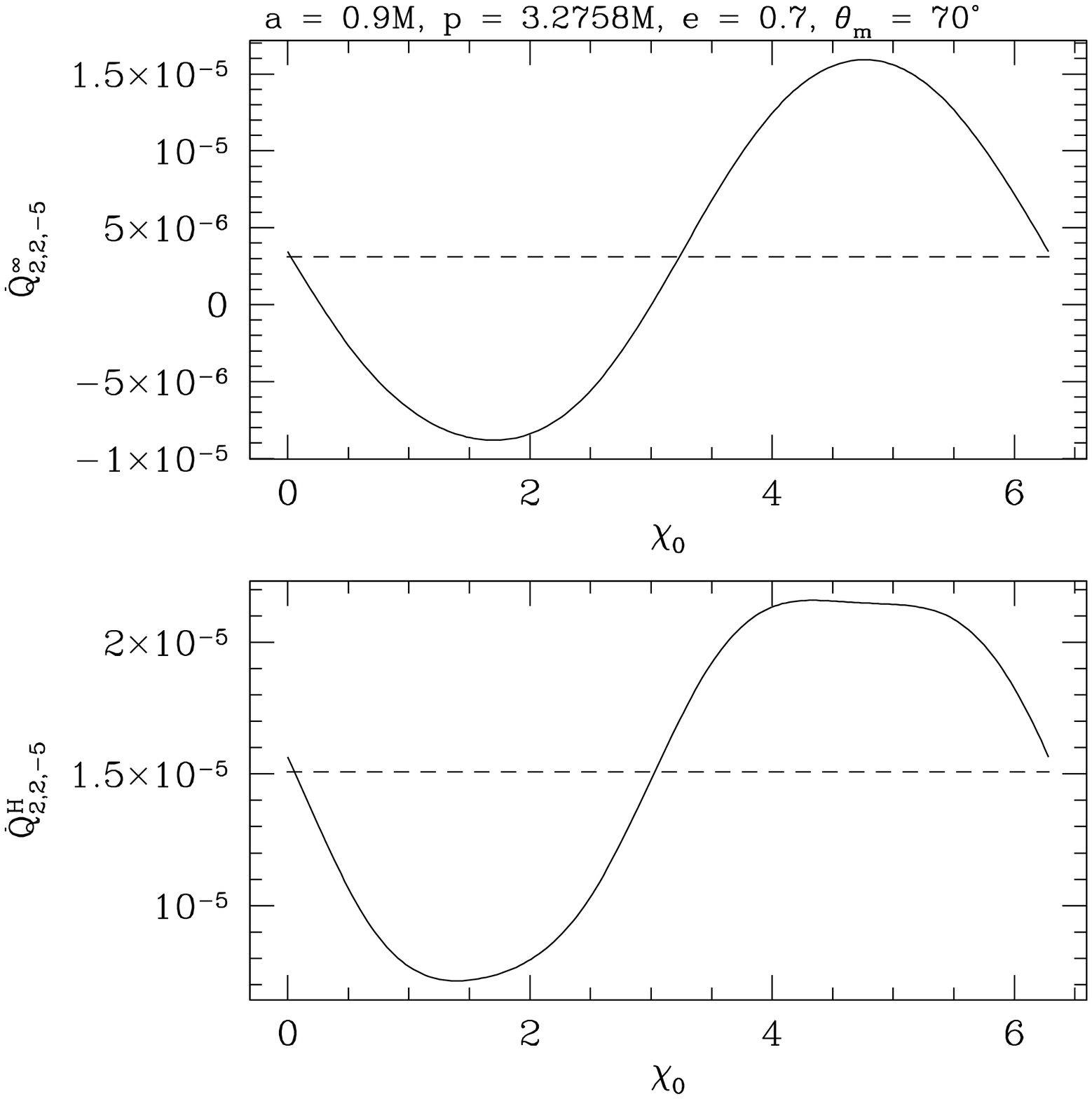}
\caption{On-resonance variation of the rates of change of orbital
  energy (left panels) and Carter constant (right panels) in the $l =
  2$, $m = 2$, $N = -5$ mode for an orbit with $p = 3.2758M$, $e =
  0.7$, $\theta_m = 70^\circ$, $a = 0.9M$ (for which
  $\Omega_\theta/\Omega_r = 3$).  Top panels give the flux to
  infinity, bottom ones give flux down the horizon.  The dashed line
  in all panels shows the value that would be obtained if the
  resonance were neglected (i.e., simply adding in quadrature the
  various 4-index amplitudes $Z^\star_{lmkn}$ that contribute to the
  3-index amplitude ${\cal Z}^\star_{lmN}$).  In all cases, the flux
  varies considerably with the phase $\chi_0$.  The variation in $\dot
  Q^\infty$ is especially interesting in this case, changing sign at
  $\chi_0 \simeq 0.28$ and $\chi_0 \simeq 3.01$.  We do not show $\dot
  L^\star_z(\chi_0)$ for this mode, since it is identical to $\dot
  E^\star(\chi_0)$ modulo a factor of $m/\omega_{mN}$.}
\label{fig:EdotQdot31}
\end{figure*}

\begin{figure*}[ht]
\includegraphics[width = 0.48\textwidth]{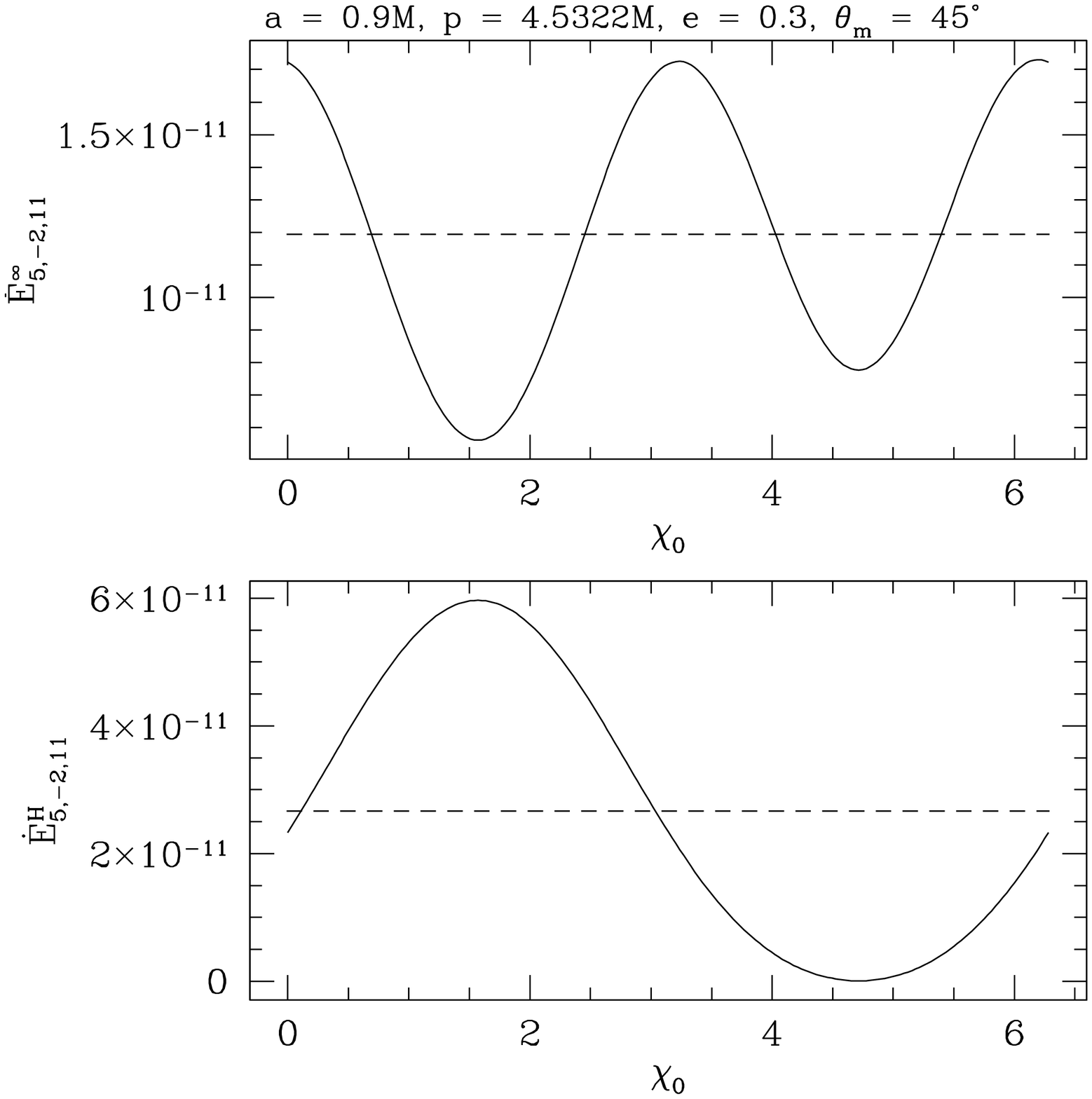}
\includegraphics[width = 0.48\textwidth]{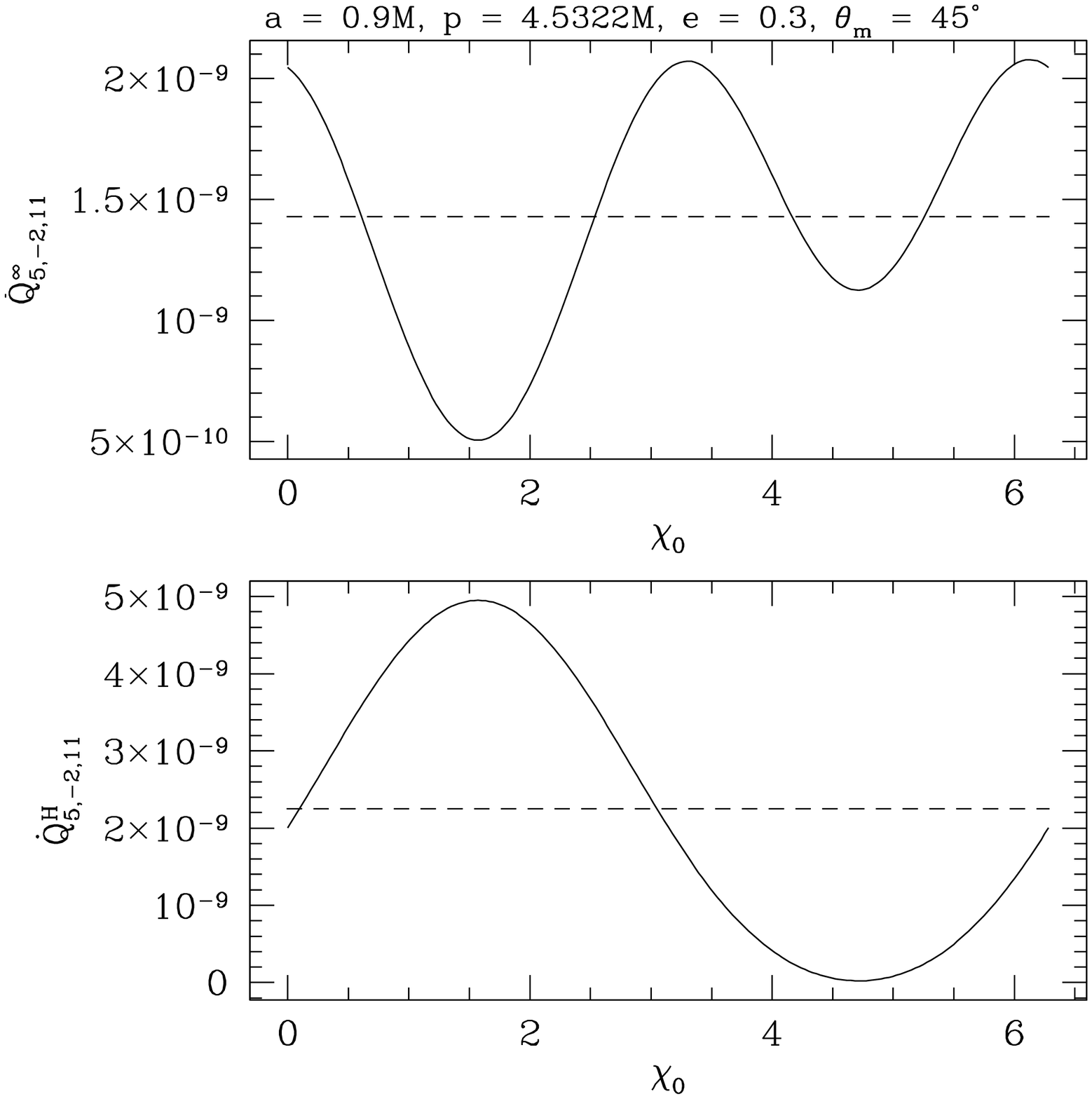}
\caption{On-resonance variation of the rates of change of orbital
  energy (left panels) and Carter constant (right panels) in the $l =
  5$, $m = -2$, $N = 11$ mode for an orbit with $p = 4.5322M$, $e =
  0.3$, $\theta_m = 45^\circ$, $a = 0.9M$ (for which
  $\Omega_\theta/\Omega_r = 2$).  Top panels give the flux to
  infinity, bottom ones give flux down the horizon.  The dashed line
  gives the value found when the resonance is neglected.  As in
  Fig.\ {\ref{fig:EdotQdot31}}, we see that $\dot E^\star_{lmN}$ and
  $\dot Q^\star_{lmN}$ vary quite a bit as $\chi_0$ sweeps from $0$ to
  $2\pi$, with minima near zero in this case for the down-horizon
  quantities.}
\label{fig:EdotQdot21}
\end{figure*}

Figures {\ref{fig:EdotQdot31}} and {\ref{fig:EdotQdot21}} give two
examples of the on-resonance rate of change of orbital constants.  We
show $\dot E^\star_{lmN}$ and $\dot Q^\star_{lmN}$ for two orbits
about a black hole with $a = 0.9$.  Figure {\ref{fig:EdotQdot31}}
shows the $l = 2$, $m = 2$, $N = -5$ mode computed for an orbit with
$p = 3.2758M$, $e = 0.7$, $\theta_m = 70^\circ$; in this case,
$\Omega_\theta/\Omega_r = 3$.  Figure {\ref{fig:EdotQdot21}} shows the
$l = 5$, $m = -2$, $N = 11$ mode for an orbit with $p = 4.5322M$, $e =
0.3$, $\theta_m = 45^\circ$, for which $\Omega_\theta/\Omega_r = 2$.

In both cases, the flux of energy to infinity varies by a factor of
about $3.1$.  This agreement is a coincidence.  The down-horizon flux
shows more variety, varying by a factor of about $6.8$ for the 3:1
resonance, and by a factor of nearly $10^3$ for the 2:1 case.  (This
large variation is because the flux comes close to zero at $\chi_0
\simeq 4.7$.)  The variation in $\dot Q^\infty_{2,2,-5}$ is especially
interesting for the 3:1 resonance: It is negative over nearly half the
span of $\chi_0$, but is positive elsewhere.  This behavior is unique
to the on-resonance form of $\dot Q^\star_{lmN}$, and arises from the
fact that it contains a term proportional to ${\rm Re}\left[{\cal
    Z}_{lmN}{\bar{\cal Y}}_{lmN}\right]$.  Because the amplitudes
${\cal Z}_{lmN}$ and ${\cal Y}_{lmN}$ can have different phases, the
behavior of $\dot Q^\star_{lmN}$ can be more complicated than the
behavior of the energy or angular momentum fluxes.  Those fluxes are
both proportional to $|{\cal Z}^\star_{lmN}|^2$, and hence are
positive or negative definite.

The horizontal dashed lines in these figures gives the rate of change
that would be found if the resonance were neglected.  In other words,
it shows the rate of change one would find by simply combining in
quadrature all of the 4-index amplitudes $Z^\star_{lmkn}$ which
contribute to the relevant 3-index amplitude ${\cal
  Z}^\star_{lmN}(\chi_0)$.  Its value is the average with respect to
$\lambda^\theta_0$ of the resonant
flux:
\begin{eqnarray}
\dot E^{\star,\ {\rm no-res}}_{lmN} &=& \frac{\Upsilon_\theta}{2\pi}
\int_0^{2\pi/\Upsilon_\theta}\dot E^\star_{lmN}\, d\lambda^\theta_0 \nonumber \\
&=& \frac{\Upsilon_\theta}{2\pi}
\int_0^{2\pi}\dot E^\star_{lmN}(\chi_0)
\frac{d\lambda^\theta_0}{d\chi_0}\,d\chi_0\;.
\label{eq:averes}
\end{eqnarray}
Recall that the parameter $\lambda^\theta_0$, introduced in
Eq.\ (\ref{eq:xi_val}), sets the value of $\lambda^\theta$ at which
$\theta = \theta_m$.  An explicit expression for the Jacobian
$d\lambda_0^\theta/d\chi_0$ is given in Eq.\ (3.76) of
Ref.\ \cite{dfh05}.  It is not difficult to show that this result must
hold\footnote{At one point in our analysis, preliminary results
  indicated that averages did not respect Eq.\ (\ref{eq:averes}).
  Gabriel Perez-Giz insisted to one of us (SAH) that this must be an
  error.  Indeed, these preliminary results were wrong.}: combining
Eqs.\ (\ref{eq:Zres_expand}) and (\ref{eq:edotInf_res}), we have
\begin{eqnarray}
\dot E^\infty_{lmN}(\chi_0) &=&
\frac{1}{4\pi\omega_{mN}^2}\left(\sum |\check{Z}^H_{lmkn}|^2
\right.\nonumber\\
&+& \left.  \sum \check{Z}^H_{lmkn}{\bar{\check{Z}}}^H_{lmk'n'}
e^{i\left[\xi_{mkn}(\chi_0) -\xi_{mk'n'}(\chi_0)\right]}\right)\;.
\nonumber\\
\end{eqnarray}
The first sum in this expression is, as usual, taken over all pairs
$(k,n)_N$, as defined earlier.  The second sum is taken over the pair
of pairs $(k,n)_N$ and $(k',n')_N$, with $k \ne k'$, $n \ne n'$.  The
first sum is exactly $\dot E^{\infty,\ {\rm no-res}}_{lmN}$.  Using
Eq.\ (\ref{eq:xi_val}), we see that on resonance,
\begin{equation}
\xi_{mkn} - \xi_{mk'n'} = (k-k')\Upsilon^\theta\lambda^\theta_0\;.
\end{equation}
Hence this term averages to zero, demonstrating the validity of
Eq.\ (\ref{eq:averes}).  Similar results hold for all of the other
rates of change we compute in this paper.  An alternative
demonstration of the identity (\ref{eq:averes}) in a more general
context can be found in Appendix C2 of Ref.\ \cite{glpgII}.

These examples show that the flux carried in each mode can vary
significantly as a function of $\chi_0$.  This shows that in principle
resonances can have a strong impact on gravitational-wave fluxes.
Notice, though, that the detailed dependence of each mode on $\chi_0$
varies quite a bit from mode to mode.  It would not be surprising if
much of the variation cancels out after summing over many modes.  We
examine this in the next section, checking to see how much flux
variation remains when many modes are added.

\subsection{Sum over many modes: Variation of total flux}
\label{sec:totalflux}

We now examine the variation in total flux on resonant orbits,
computing the sums (\ref{eq:edotInf_res}) and (\ref{eq:edotH_res}).
Those sums are taken over an infinite number of modes, which we cannot
do in a numerical calculation.  We instead truncate the sum over index
$l$ at $l_{\rm max} = 6$; for orbits with $e = 0.3$, we truncate the
sum over $N$ at $N_{\rm max} = 50$, and truncate at $N_{\rm max} =
100$ for $e = 0.7$:
\begin{equation}
\dot E^\star(\chi_0) = \sum_{l = 2}^{l_{\rm max}}\sum_{m =
  -l}^l\sum_{N = -N_{\rm max}}^{N_{\rm max}} \dot
E^\star_{lmN}(\chi_0)\;.
\label{eq:p_to_t}
\end{equation}
We have not performed a careful convergence analysis, but have found
that increasing $l_{\rm max}$ and $N_{\rm max}$ only changes our
numerical results by an unimportant fraction for the orbits we have
examined so far.  We do not claim our accuracy to be good enough for
``production'' purposes, but claim it is good enough to illustrate the
physics that we present here.

\begin{figure}
\includegraphics[width = 0.48\textwidth]{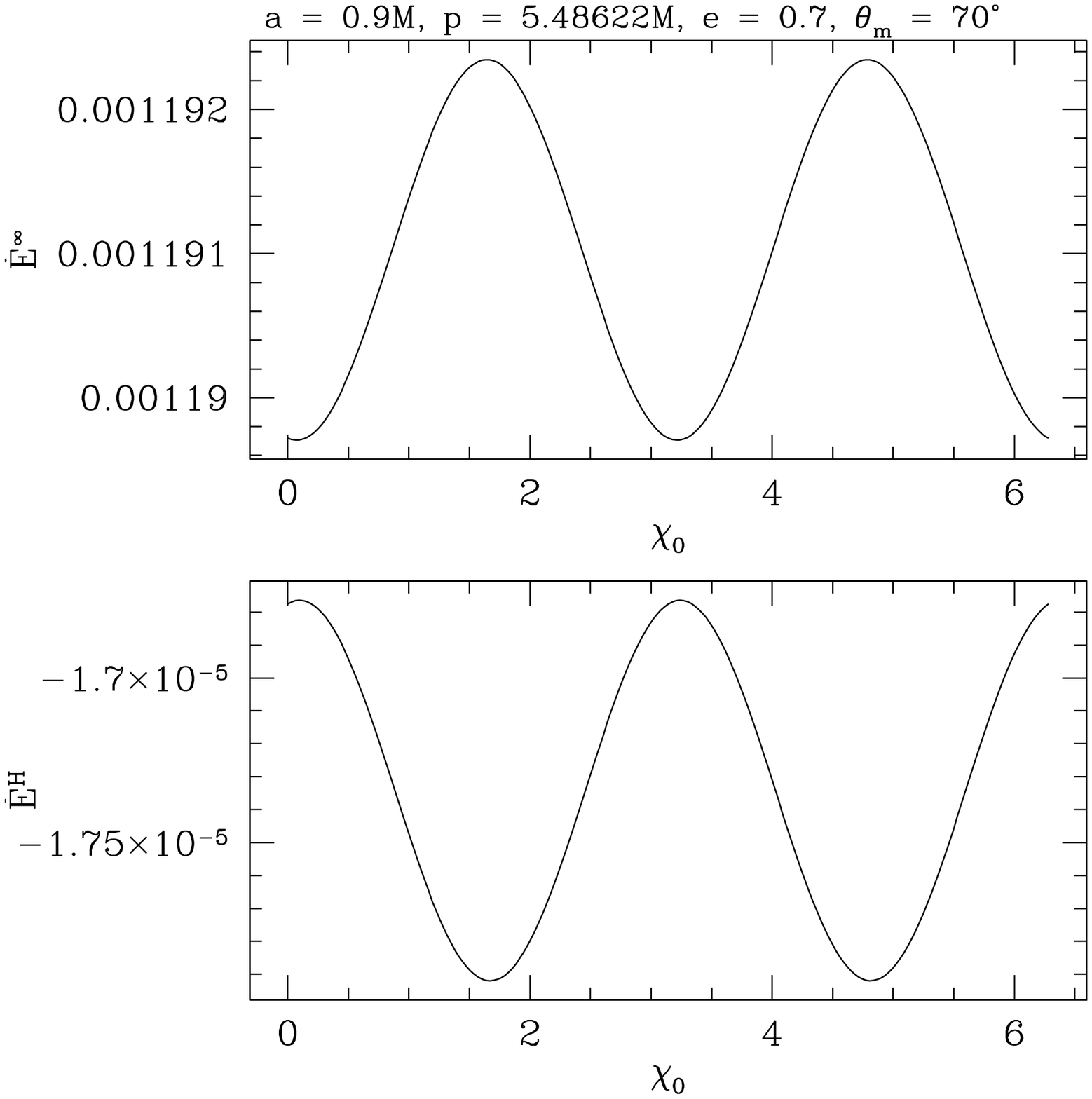}
\caption{Variation of total energy flux, both to infinity (top) and
  down the horizon (bottom) for an orbit with $p = 5.48622M$, $e =
  0.7$, $\theta_m = 70^\circ$, $a = 0.9M$ (for which
  $\Omega_\theta/\Omega_r = 3/2$).  After summing over many modes, the
  variation is significantly reduced: the flux to infinity only varies
  by about $0.127\%$, and that down the horizon varies by roughly
  $1.6\%$.  The variations in $\dot L_z^\star(\chi_0)$ and $\dot
  Q^\star(\chi_0)$ are qualitatively similar, so we do not show them.
  See Table {\ref{tab:variations_e0.7_thm70}} for more details.}
\label{fig:Edottot31}
\end{figure}

Figure {\ref{fig:Edottot31}} shows one example of how, after summing
over many modes, $\dot E^\star$ varies as a function of $\chi_0$.  We
put $a = 0.9M$, and choose an orbit with $p = 5.48622M$, $e = 0.7$,
and $\theta_m = 70^\circ$, for which $\Omega_\theta/\Omega_r = 3/2$.
The fractional variation in $\dot E^\star$ we find is much smaller
than the variation we saw in individual modes: the summed flux to
infinity varies by about $0.2\%$, and the down-horizon flux varies by
about $6.7\%$.  The down-horizon flux is much smaller than the flux to
infinity, so the variations are dominated by the behavior of $\dot
E^\infty$.  The behaviors of $\dot L_z^\star(\chi_0)$ and $\dot
Q^\star(\chi_0)$ are qualitatively similar to $\dot E^\star(\chi_0)$,
so we do not show plots for those quantities.

Tables {\ref{tab:variations_e0.7_thm20}} --
{\ref{tab:variations_e0.3_thm70}} present the fractional variation in
$\dot E^\star$, $\dot L_z^\star$, and $\dot Q^\star$ for several
orbits about a black hole with spin $a = 0.9M$.  Within each table, we
fix $e$ and $\theta_m$.  We look at large and small eccentricity ($e =
0.7$ and $e = 0.3$), and large and small orbital
inclination\footnote{Note that smaller $\theta_m$ implies a more
  highly inclined orbit; $\theta_m = 90^\circ$ is an equatorial
  orbit.} ($\theta_m = 20^\circ$ and $\theta_m = 70^\circ$).  We then
vary $p$ to study radiation emission from four different resonances,
3:1, 2:1, 3:2, and 4:3.  The fractional variation in a quantity $X$ is
defined as
\begin{equation}
\Delta X \equiv \frac{|X_{\rm max}| - |X_{\rm min}|}{(|X_{\rm max}| +
  |X_{\rm min}|)/2}\;,
\label{eq:DeltaX}
\end{equation}
where $X_{\rm max/min}$ is the maximum or minimum value $X$ takes as
$\chi_0$ varies from $0$ to $2\pi$.

The peak-to-trough variation (\ref{eq:DeltaX}) in the fluxes is an
important quantity that determines several properties of the
resonances.  First, the ``kicks'' in $E$, $L_z$, and $Q$ that occur as
a system spirals through a resonance are directly proportional to the
variation (\ref{eq:DeltaX}) {\cite{fhhr_inprep}}.  As such, these
quantities give some idea of how much impact resonances will have as a
system evolves through orbit, even though we have not yet developed
the tools needed to compute these evolutions in detail.  Second, there
are two qualitatively different types of resonances that can occur in
systems of this kind: a simple linear resonance in which the kicks
depend sinusoidally on the phase parameter $\chi_0$ (cf.\ the final
equation of FH), and a nonlinear variant in which the dynamics is
rather more complicated.  For the nonlinear scenario, it is possible
to have a ``sustained resonance'' in which the system becomes trapped
near the resonance for an extended period of time
\cite{Ori,Kevorkian}.  Our numerical results show that $\Delta X \ll
1$ at least over all of the parameter space we have surveyed so far,
which indicates that the resonances are always of the simple, linear
kind.  This agrees with post-Newtonian analyses {\cite{fhhr_inprep}},
as well as recent work by van de Meent {\cite{vdM}}.

\begin{table*}
\begin{ruledtabular}
\begin{tabular}{|c|c|c|c||c|c|c||c|c|c||c|c|c|}
$e$ & $\theta_m$ &
$p$ &$\Omega_\theta/\Omega_r$ & $\Delta\dot E^H$ & $\Delta\dot L_z^H$ & $\Delta\dot Q^H$ & $\Delta\dot E^\infty$ & $\Delta\dot L_z^\infty$ & $\Delta\dot Q^\infty$ & $\Delta\dot E^{\rm tot}$ & $\Delta\dot L_z^{\rm tot}$ & $\Delta\dot Q^{\rm tot}$ \\
\hline
$0.7$ & $20^\circ$ &
$5.38952M$ & $3$ & $92.5\%$ & $0.363\%$ & $0.543\%$ & $0.087\%$ & $0.069\%$ & $0.105\%$ & $0.125\%$ & $0.027\%$ & $0.126\%$
\\
\hline
$0.7$ & $20^\circ$ &
$6.31541M$ & $2$ & $30.7\%$ & $2.89\%$ & $1.82\%$ & $0.634\%$ & $0.483\%$ & $0.467\%$ & $0.662\%$ & $0.270\%$ & $0.494\%$
 \\
\hline
$0.7$ & $20^\circ$ &
$8.77436M$ & $3/2$ & $106\%$ & $21.9\%$ & $10.4\%$ & $1.17\%$ & $0.172\%$ & $0.219\%$ & $1.03\%$ & $0.489\%$ & $0.261\%$
 \\
\hline
$0.7$ & $20^\circ$ &
$11.4219M$ & $4/3$ & $1.41\%$ & $0.117\%$ & $0.979\%$ & $0.048\%$ & $0.058\%$ & $0.003\%$ & $0.047\%$ & $0.060\%$ & $0.002\%$
 \\
\end{tabular}
\end{ruledtabular}
\caption{Variation in flux for orbits with $e = 0.7$ and $\theta_m =
  20^\circ$ about a black hole with spin $a = 0.9M$.  We vary $p$ to
  examine a sequence of orbital resonances from
  $\Omega_\theta/\Omega_r = 3$ to $\Omega_\theta/\Omega_r = 4/3$.
  Columns 3 -- 5 show the fractional variation in energy flux, axial
  angular momentum flux, and Carter constant rate of change arising
  from the down-hole fields; the fractional variation is defined
  precisely in the text.  Columns 6 -- 8 repeat this information for
  these fields at infinity, and columns 9 -- 11 give the fractional
  variation for the totals (infinity plus horizon).  The variations
  are largest for the 3:2 resonance and 2:1 resonances (depending on
  which quantity we examine), and smallest for the 4:3 resonance.}
\label{tab:variations_e0.7_thm20}
\end{table*}

\begin{table*}
\begin{ruledtabular}
\begin{tabular}{|c|c|c|c||c|c|c||c|c|c||c|c|c|}
$e$ & $\theta_m$ &
$p$ &$\Omega_\theta/\Omega_r$ & $\Delta\dot E^H$ & $\Delta\dot L_z^H$ & $\Delta\dot Q^H$ & $\Delta\dot E^\infty$ & $\Delta\dot L_z^\infty$ & $\Delta\dot Q^\infty$ & $\Delta\dot E^{\rm tot}$ & $\Delta\dot L_z^{\rm tot}$ & $\Delta\dot Q^{\rm tot}$ \\
\hline
$0.7$ & $70^\circ$ &
$3.27580M$ & $3$ & $1.14\%$ & $1.89\%$ & $2.60\%$ & $0.010\%$ & $0.067\%$ & $0.421\%$ & $0.026\%$ & $0.009\%$ & $0.035\%$
\\
\hline
$0.7$ & $70^\circ$ &
$3.78947M$ & $2$ & $1.60\%$ & $2.68\%$ & $6.01\%$ & $0.204\%$ & $0.153\%$ & $0.109\%$ & $0.167\%$ & $0.067\%$ & $0.357\%$
 \\
\hline
$0.7$ & $70^\circ$ &
$5.48622M$ & $3/2$ & $6.66\%$ & $5.77\%$ & $26.3\%$ & $0.222\%$ & $0.034\%$ & $0.216\%$ & $0.127\%$ & $0.078\%$ & $0.210\%$
 \\
\hline
$0.7$ & $70^\circ$ &
$7.53814M$ & $4/3$ & $0.042\%$ & $0.008\%$ & $4.04\%$ & $0.001\%$ & $0.002\%$ & $0.023\%$ & $0.001\%$ & $0.002\%$ & $0.023\%$
 \\
\end{tabular}
\end{ruledtabular}
\caption{Variation in flux for orbits with $e = 0.7$ and $\theta_m =
  70^\circ$ about a black hole with spin $a = 0.9M$.  As when $e =
  0.7$ and $\theta_m = 70^\circ$, the variations are largest for the
  3:2 resonance and 2:1 resonances (depending on which quantity we
  examine), and smallest for the 4:3 resonance.}
\label{tab:variations_e0.7_thm70}
\end{table*}

\begin{table*}
\begin{ruledtabular}
\begin{tabular}{|c|c|c|c||c|c|c||c|c|c||c|c|c|}
$e$ & $\theta_m$ &
$p$ &$\Omega_\theta/\Omega_r$ & $\Delta\dot E^H$ & $\Delta\dot L_z^H$ & $\Delta\dot Q^H$ & $\Delta\dot E^\infty$ & $\Delta\dot L_z^\infty$ & $\Delta\dot Q^\infty$ & $\Delta\dot E^{\rm tot}$ & $\Delta\dot L_z^{\rm tot}$ & $\Delta\dot Q^{\rm tot}$ \\
\hline
$0.3$ & $20^\circ$ &
$5.04884M$ & $3$ & $4.43\%$ & $0.659\%$ & $1.15\%$ & $0.027\%$ & $0.068\%$ & $0.054\%$ & $0.008\%$ & $0.024\%$ & $0.033\%$
\\
\hline
$0.3$ & $20^\circ$ &
$6.12789M$ & $2$ & $4.24\%$ & $1.42\%$ & $1.94\%$ & $0.012\%$ & $0.025\%$ & $0.013\%$ & $0.004\%$ & $0.080\%$ & $0.002\%$
 \\
\hline
$0.3$ & $20^\circ$ &
$8.65334M$ & $3/2$ & $3.34\%$ & $2.62\%$ & $8.82\%$ & $0.308\%$ & $0.158\%$ & $0.114\%$ & $0.303\%$ & $0.123\%$ & $0.123\%$
 \\
\hline
$0.3$ & $20^\circ$ &
$11.3158M$ & $4/3$ & $0.104\%$ & $0.165\%$ & $1.09\%$ & $0.003\%$ & $0.005\%$ & $0.002\%$ & $0.003\%$ & $0.004\%$ & $0.002\%$
 \\
\end{tabular}
\end{ruledtabular}
\caption{Variation in flux for orbits with $e = 0.3$ and $\theta_m =
  20^\circ$ about a black hole with spin $a = 0.9M$.  In this case,
  the 3:2 resonance shows larger variations than all other cases; the
  2:1 resonance is surprisingly weak, given its strength in other
  examples we have seen.  As usual, however, the 4:3 resonance shows
  the least amount of variation among all the resonances that we
  consider.}
\label{tab:variations_e0.3_thm20}
\end{table*}

\begin{table*}
\begin{ruledtabular}
\begin{tabular}{|c|c|c|c||c|c|c||c|c|c||c|c|c|}
$e$ & $\theta_m$ &
$p$ &$\Omega_\theta/\Omega_r$ & $\Delta\dot E^H$ & $\Delta\dot L_z^H$ & $\Delta\dot Q^H$ & $\Delta\dot E^\infty$ & $\Delta\dot L_z^\infty$ & $\Delta\dot Q^\infty$ & $\Delta\dot E^{\rm tot}$ & $\Delta\dot L_z^{\rm tot}$ & $\Delta\dot Q^{\rm tot}$ \\
\hline
$0.3$ & $70^\circ$ &
$2.91117M$ & $3$ & $1.13\%$ & $1.17\%$ & $0.544\%$ & $0.023\%$ & $0.026\%$ & $0.367\%$ & $0.059\%$ & $0.070\%$ & $0.310\%$
\\
\hline
$0.3$ & $70^\circ$ &
$3.55601M$ & $2$ & $1.10\%$ & $1.28\%$ & $3.67\%$ & $0.103\%$ & $0.142\%$ & $0.039\%$ & $0.131\%$ & $0.179\%$ & $0.046\%$
 \\
\hline
$0.3$ & $70^\circ$ &
$5.34138M$ & $3/2$ & $0.481\%$ & $0.336\%$ & $4.86\%$ & $0.106\%$ & $0.063\%$ & $0.227\%$ & $0.102\%$ & $0.067\%$ & $0.208\%$
 \\
\hline
$0.3$ & $70^\circ$ &
$7.41979M$ & $4/3$ & $0.007\%$ & $0.021\%$ & $0.229\%$ & $0.001\%$ & $0.001\%$ & $0.006\%$ & $0.001\%$ & $0.001\%$ & $0.006\%$
 \\
\end{tabular}
\end{ruledtabular}
\caption{Variation in flux for orbits with $e = 0.3$ and $\theta_m =
  70^\circ$ about a black hole with spin $a = 0.9M$.  The case is
  qualitatively similar to most of the others, with the 3:2 and 2:1
  showing the largest degree of variation (depending on the quantity
  being examined), and the 4:3 case showing the least.}
\label{tab:variations_e0.3_thm70}
\end{table*}

Some interesting trends are apparent from these tables.  First, notice
that in all cases the down-horizon variation is quite a bit larger
than than the variation in the quantities to infinity.  However, in
all cases, the magnitude of the down-horizon fluxes is substantially
smaller than the magnitude to infinity.  The total variations are thus
dominated by the fluxes to infinity, consistent with the results shown
in Fig.\ {\ref{fig:Edottot31}}.

\begin{figure}
\includegraphics[width = 0.48\textwidth]{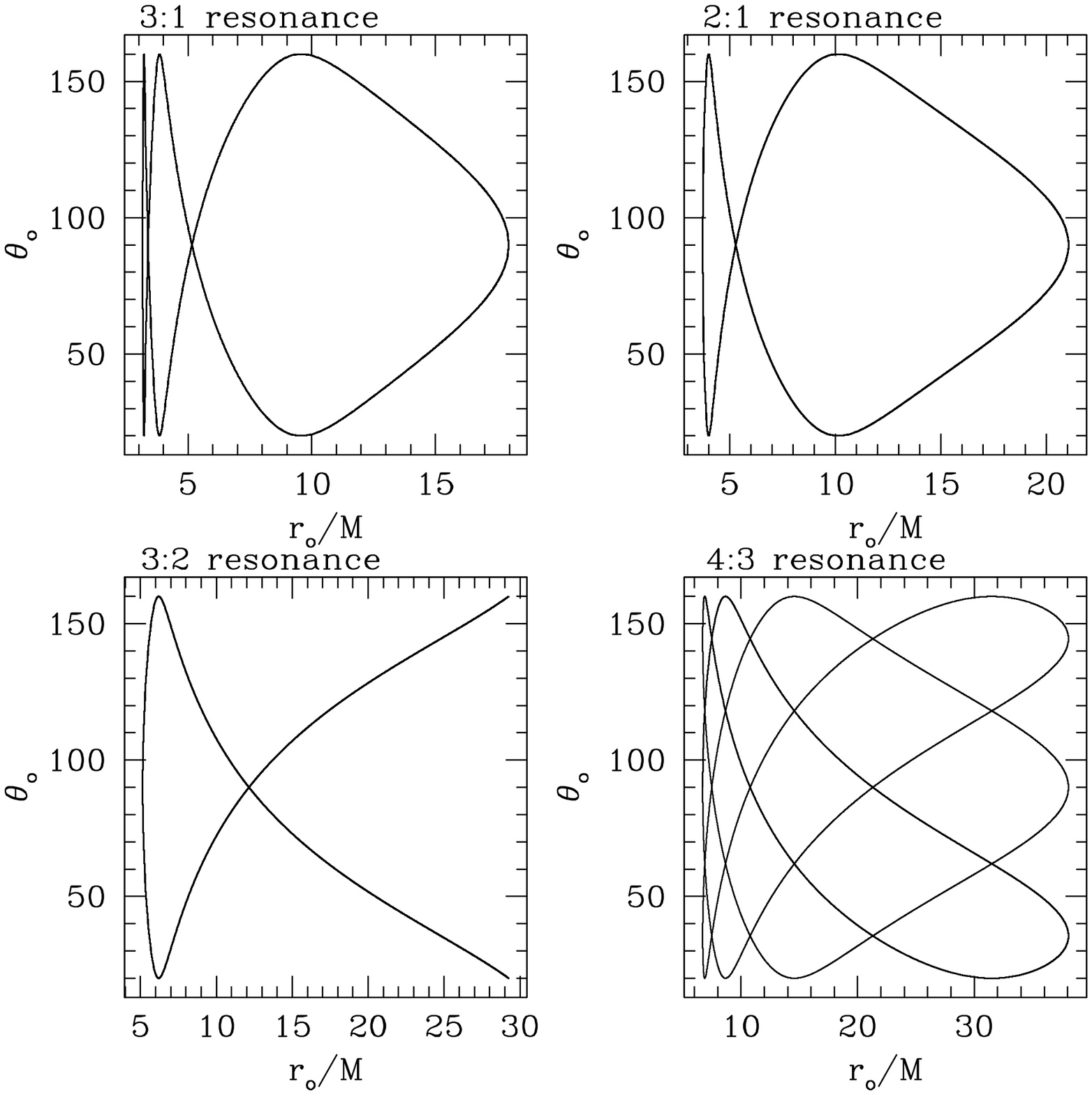}
\caption{Trajectories in the $(r,\theta)$ plane for the orbits
  discussed in Table {\ref{tab:variations_e0.7_thm20}}.  We put
  $\chi_0 = \pi/2$ for these plots.  The 4:3 resonance shows the
  smallest flux variation of those considered here, and has the most
  complicated trajectory.  This orbit comes ``close to'' enough points
  in the $(r,\theta)$ plane that it averages over much of its
  accessible domain.  By contrast, the 3:2 and 2:1 orbits have simple
  trajectories and do not effectively average over their domain.
  Fluxes from these orbits tend to show the largest variation with
  $\chi_0$.  The 3:1 orbit is similar to the 2:1 orbit, but with an
  additional angular oscillation at small radius which enhances
  orbital averaging.  This orbit generally shows intermediate flux
  variation compared with the other cases.}
\label{fig:compareres}
\end{figure}

Second, notice that the largest variations are seen in either the 2:1
or 3:2 resonances (always the 3:2 resonance for orbits with $e = 0.3$,
but either 3:2 or 2:1 depending on which quantity we examine for the
orbits with $e = 0.7$).  The variations are consistently smallest for
the 4:3 resonance.  This behavior correlates with the shape that a
resonant orbit traces in the $(r,\theta)$ plane.  Figure
{\ref{fig:compareres}} shows these orbital tracks for the four orbits
presented in Table {\ref{tab:variations_e0.7_thm20}}.  For simplicity,
we only show tracks for $\chi_0 = \pi/2$.

The contrasting shapes of the 2:1 and 3:2 orbits on one hand, and of
the 4:3 orbit on the other, are particularly noteworthy.  The 4:3
resonant orbit (bottom right) traces a rather complicated Lissajous
figure which comes ``close to'' many of the $(r,\theta)$ points which
are accessible given $(p,e,\theta_m)$.  This complicated trajectory
samples much of the accessible domain in $r$ and $\theta$.  Appealing
to the constrained integral method of computing ${\cal Z}^\star_{lmN}$
(cf.\ Sec.\ {\ref{sec:resII}}), we can say that the motion effectively
averages out the variations in the integrand by passing close to so
many accessible points.

By contrast, the trajectory for the 2:1 and 3:2 resonances (top right
and bottom left) are much simpler.  These trajectories do not come as
close to so many points in their allowed domain, and so do not average
the variations in their integrands as effectively.  The trajectory for
the 3:1 (top left) resonance is similar to that for the 2:1 case, but
with an additional angular oscillation at small radius.  This extra
oscillation enhances the averaging as the orbit moves through a
particularly strong-field part of its domain.  Not too surprisingly,
the flux variation in this case is generally intermediate to the
others.

Beyond the fact that orbits with simple shapes in the $(r,\theta)$
plane tend to show stronger resonances than orbits with more
complicated shapes, we do not as yet see strong evidence of any trend
which would allow us to predict which resonances will tend to be
``strong'' (i.e., exhibit large variation in orbital parameter
evolution) and which ``weak.''  Consider for example the rate of
change of orbital energy, $\Delta\dot E^{\rm tot}$.  As we go from
high inclination to shallow and from high eccentricity to low, we see
that $\Delta\dot E^{\rm tot}$ goes from large to small: It takes the
value $1.03\%$ for high eccentricity, high inclination (Table
{\ref{tab:variations_e0.7_thm20}}); $0.167\%$ and $0.303\%$ for the
mixed cases (Tables {\ref{tab:variations_e0.7_thm70}} and
{\ref{tab:variations_e0.3_thm20}}); and the value $0.131\%$ for the
case of small eccentricity, shallow inclination (Table
{\ref{tab:variations_e0.3_thm70}}).  This appears to suggest, at least
roughly, that the strength of the resonance is correlated with the
degree of radial and angular motion.

However, no such pattern is seen when we examine $\Delta\dot L_z^{\rm
  tot}$ and $\Delta\dot Q^{\rm tot}$.  For $L_z$, the high
eccentricity, high inclination case again produces the largest
variation ($0.489\%$, in the 3:2 resonance of Table
{\ref{tab:variations_e0.7_thm20}}).  However, the {\it low}
eccentricity, {\it low} inclination case produces the second largest
variation ($0.123\%$, in the 3:2 resonance of Table
{\ref{tab:variations_e0.3_thm70}}).  These values of $e$ and
$\theta_m$ likewise produce the largest and second-largest variations
in the Carter constant (albeit in different resonances).

We do not yet have a compelling way to explain these trends (or lack
of trends) in the resonances' strength, so we leave this mystery to
future work.

\section{Concluding discussion and future work}
\label{sec:conclude}

In this analysis, using a Teukolsky-equation-based formalism good for
exploring radiation produced by strong-field orbits, we have confirmed
the picture that on resonance the gravitational-wave driven evolution
of a binary can depend strongly on the relative phase of radial and
angular motions.  A binary in which this relative phase has the value
$\pi/2$ as the system enters resonance may evolve quite differently
from an otherwise identical system in which this phase is $3\pi/2$
entering resonance.  A typical extreme mass-ratio binary can be
expected to pass through several orbital resonances en route to its
final coalescence.  That their evolution through each resonance
depends strongly on an ``accidental'' phase parameter has the
potential to complicate schemes for measuring gravitational waves from
these binaries.

We find that the degree of variation depends strongly upon the
topology of the orbital trajectory in the $(r,\theta)$
plane\footnote{Strictly speaking, it is a trajectory's geometry that
  matters, particularly how close the orbit comes to all accessible
  points in the $(r,\theta)$ plane.  However, its geometry is strongly
  correlated to its topology, which is an invariant property of a
  resonant orbit's frequencies \cite{glpgI}.  As such, the topology is
  a valuable way to characterize this aspect of its resonant
  behavior.}.  Of the cases we have studied in detail, the orbital
plane trajectory of resonances like $\Omega_\theta/\Omega_r = 3/2$
have a simple topology.  This trajectory does not cross itself very
often, and does not come close to many points in the plane.  Such
resonances do not effectively average out the behavior of the source
to the wave equation.  As such, if the source varies significantly
over an orbit, there can be a strong residue of this variation in the
associated radiation.  By contrast, the trajectory of resonances like
$\Omega_\theta/\Omega_r = 4/3$ has a more complicated topology,
crossing itself many times, and more completely ``covering'' the
plane.  In these cases, the orbit comes ``close to'' many of the
allowed points in the $(r,\theta)$ plane, which quite effectively
averages out the source's behavior.

Although instructive and a nice validation of our ability to examine
resonances, these results are not enough to truly assess the
importance that resonances have in a strong field analysis.  We must
be able to analyze a system as it evolves through a resonance, and
thereby integrate the full ``kicks'' in the integrals of motion $E$,
$L_z$, and $Q$ imparted to the system as it passes through resonance.
A first step in this direction has been taken by van de Meent
{\cite{vdM}}, who examines the likelihood that resonances can ``trap''
an orbit, leading to long-lived resonant waves.  Part of van de
Meent's analysis is adescription of the system's evolution as motion
through a one-dimensional effective potential.  This approach is
likely to be useful for more general analysis of resonant evolution.

For our planned work, we have begun expanding our Teukolsky code to
compute, in the frequency domain, the instantaneous components of the
dissipative or radiative piece of the self force.  Our formulation is
based in part on the discussion of
Refs.\ {\cite{mino05,tanaka06,dfh05}}, but generalized to compute the
full dissipative self force rather than its torus average\footnote{One
  might be concerned about gauge ambiguities associated with the
  gravitational self force.  As shown by Mino {\cite{mino03}}, these
  ambiguities disappear when one averages the self force's effects
  over an infinite time.  In a two-timescale expansion {\cite{FHtt}},
  such ambiguities remain, but are suppressed by the ratio of the
  timescales.}.  This will allow us to study how a real inspiral is
affected as we evolve through each resonance using results that are
good deep in the strong field.  The results shown in this paper are a
first step toward this, demonstrating that our strong-field toolkit
can be used to study resonant effects.

\acknowledgments

The code used here was developed from that used for
Ref.\ {\cite{dh06}}; we thank Steve Drasco for his contributions to
that work and the development of this code.  We thank Tanja Hinderer,
Amos Ori, Nicol\'as Yunes, and Janna Levin for useful discussions in
the course of this research, and Leor Barack and Maarten van de Meent
for pointing out a minor error regarding the magnitude of orbital
frequencies for generic orbits.  We particularly thank Gabriel
Perez-Giz, whose discussion of how the on-resonant fluxes should
average away was helpful in tracking down a bug in our analysis, and
Rebecca Grossman for detailed and helpful discussions regarding
Ref.\ {\cite{glpgII}}.  We also thank this paper's referees, whose
feedback enabled us to significantly improve presentation of our
results.  Our work was supported at MIT by NSF grant PHY-1068720 and
by NASA Grant NNX08AL42G, and at Cornell by NSF grant PHY-1068541.
SAH gratefully acknowledges fellowship support by the John Simon
Guggenheim Memorial Foundation, and sabbatical support from the
Canadian Institute for Theoretical Astrophysics and the Perimeter
Institute for Theoretical Physics.  UR gratefully acknowledges
fellowship support from the Royal Thai Government.

\appendix

\section{Proof: Equivalence of methods for computing on-resonant
amplitudes}
\label{app:equivalence}

In this appendix, we prove that Eq.\ (\ref{eq:Zres_int}), the 1-D
integral for the on-resonance 3-index amplitude ${\cal
  Z}^\star_{lmN}(\chi_0)$, is equivalent to
Eq.\ (\ref{eq:Zres_expand}), the on-resonance amplitude expressed as a
sum of 4-index amplitudes $Z^\star_{lmkn}(\chi_0)$, each of which is
computed using the 2-D integral (\ref{eq:Zcoef}).  Similar discussion,
demonstrating the equivalence of these forms of the amplitudes, can be
found in Appendix B of Ref.\ {\cite{glpgII}}.

We begin with Eq.\ (\ref{eq:Z2}), which we repeat here:
\begin{equation}
Z^\star_{lm\omega}(\chi_0) = C^\star\!\!\int_{-\infty}^\infty
\!\!\!
d\lambda\,e^{i(\omega\Gamma - m\Upsilon_\phi)\lambda}
J^\star_{lm\omega}[r_{\rm o}(\lambda),\theta_{\rm o}(\lambda,\chi_0)]\;.
\end{equation}
Recall that the ``o'' subscript on $r$ and $\theta$ means that those
are quantities along the orbit, and as such vary at harmonics of the
frequencies $\Upsilon_r$ and $\Upsilon_\theta$.  We can thus expand
$J^\star_{lm\omega}$ in a Fourier series:
\begin{eqnarray}
J^\star_{lm\omega} &=& \sum_{kn} J^\star_{\omega lmkn}(\chi_0)
e^{-i(k\Upsilon_\theta + n\Upsilon_r)\lambda}\;,
\label{eq:Jstar_gen}\\
&=& \sum_{N} {\cal J}^\star_{\omega lmN}(\chi_0)
e^{-iN\Upsilon\lambda}\;.
\label{eq:Jstar_res}
\end{eqnarray}
Equation (\ref{eq:Jstar_gen}) holds for arbitrary orbits.  Equation
(\ref{eq:Jstar_res}) only holds on resonance, when $\Upsilon_\theta =
\beta_\theta\Upsilon$, $\Upsilon_r = \beta_r\Upsilon$.

Because Eq.\ (\ref{eq:Jstar_gen}) remains valid for resonant orbits,
in the resonant case
\begin{equation}
\sum_{N} {\cal J}^\star_{\omega lmN}(\chi_0) e^{-iN\Upsilon\lambda}
\doteq \sum_{kn} J^\star_{\omega lmkn}(\chi_0) e^{-i(k\Upsilon_\theta
  + n\Upsilon_r)\lambda}\;.
\label{eq:res_condition}
\end{equation}
(The notation ``$\doteq$'' means that this equation is true only on
resonance.)  Multiply both sides by $e^{iN'\Upsilon\lambda}$ and
integrate from $0$ to $2\pi/\Upsilon$.  On the left-hand side, we have
\begin{eqnarray}
& &\int_0^{2\pi/\Upsilon} \sum_N {\cal J}^\star_{\omega lmN}(\chi_0)
e^{i(N'-N)\Upsilon\lambda}d\lambda\qquad\qquad\qquad
\nonumber\\
& &\qquad\qquad\qquad\qquad
 = \frac{2\pi}{\Upsilon}\sum_N
{\cal J}^\star_{\omega lmN}(\chi_0) \delta_{NN'}
\nonumber\\
& &\qquad\qquad\qquad\qquad
 = \frac{2\pi}{\Upsilon}
{\cal J}^\star_{\omega lmN'}(\chi_0)\;.
\label{eq:proof_lhs}
\end{eqnarray}
To do this operation on the right-hand side, first note that
by the resonance condition we must have
\begin{equation}
k\Upsilon_\theta + n\Upsilon_r = (k\beta_\theta + n\beta_r)\Upsilon\;.
\end{equation}
Using this, the integral for the right-hand side becomes
\begin{eqnarray}
& &\int_0^{2\pi/\Upsilon} \sum_{kn} J^\star_{\omega lmkn}(\chi_0)
e^{i[N'-(k\beta_\theta + n\beta_r)]\Upsilon\lambda}d\lambda\qquad\qquad
\nonumber\\
& &\qquad\qquad\qquad = \frac{2\pi}{\Upsilon}\sum_{kn}
J^\star_{\omega lmkn}(\chi_0) \delta_{(k\beta_\theta + n\beta_r),N'}
\nonumber\\
& &\qquad\qquad\qquad = \frac{2\pi}{\Upsilon}\sum_{(k,n)_{N'}}
J^\star_{\omega lmkn}(\chi_0)\;.
\label{eq:proof_rhs}
\end{eqnarray}
The notation $(k,n)_{N'}$ means that the sum is over all pairs $(k,n)$
which satisfy $k\beta_\theta + n\beta_r = N'$.

Next, use Eqs.\ (\ref{eq:Zcoef1}), (\ref{eq:Zcoefa}) and
(\ref{eq:Zres_int1}), invoke
Eq.\ (\ref{eq:Zres_expand}), drop the primes on the index $N$, and
equate (\ref{eq:proof_lhs}) and (\ref{eq:proof_rhs}).  The result is
\begin{equation}
{\cal Z}^\star_{lmN}(\chi_0) \doteq \sum_{(k,n)_N}
e^{i\xi_{mkn}(\chi_0)} \check{Z}^\star_{lmkn}\;,
\end{equation}
which proves that the 1-D integral and the sum of 2-D integrals are
equivalent for resonant orbits.

\section{Evolution of the Carter constant}
\label{app:qdot}

The third conserved quantity associated with orbits of Kerr black
holes is the Carter constant, $Q$.  Rearranging
Eq.\ (\ref{eq:thetadot}), we write
\begin{equation}
Q = \cot^2\theta\,L_z^2 + \cos^2\theta(1 - E^2) +
\left(\frac{d\theta}{d\lambda}\right)^2\;.
\end{equation}
Reference {\cite{sago}} (S06) first demonstrated how to compute the
long-time-averaged evolution of $Q$, at least for non-resonant orbits.
In this appendix, we revisit their calculation in some detail in order
to see clearly how it will have to be modified for resonant orbits
(modifying some details to be in accord with our notation).  We then
examine how the calculation changes when we are on an orbital
resonance.

\subsection{A comment regarding averaging}
\label{sec:averaging}

In this and the following appendix, we average several quantities,
defining
\begin{equation}
\langle f\rangle = \lim_{L \to \infty} \frac{1}{2L}\int_{-L}^L
d\lambda\,f(\lambda)
\label{eq:avedef}
\end{equation}
for various functions $f = f[r(\lambda),\theta(\lambda)]$.  For
non-resonant orbits (i.e, those in which $\Omega_\theta/\Omega_r$ is
an irrational number), the average (\ref{eq:avedef}) is equivalent to
the torus average:
\begin{equation}
\langle f\rangle_{\rm non-res} = \frac{\Upsilon_\theta\Upsilon_r}{(2\pi)^2}
\int_0^{2\pi/\Upsilon_\theta}\!\!\! \int_0^{2\pi/\Upsilon_r}\!\!\!\!
f[r(\lambda^r),\theta(\lambda^\theta)]\,d\lambda^r\,d\lambda^\theta\;.
\label{eq:torusave2D}
\end{equation}
If the orbit's frequencies are commensurate (i.e., if it is a resonant
orbit), (\ref{eq:avedef}) is equivalent to the average over the 1-D
trajectory that the orbit traces on the $(\lambda^r,\lambda^\theta)$
torus:
\begin{equation}
\langle f\rangle_{\rm res} = \frac{\Upsilon}{2\pi}
\int_0^{2\pi/\Upsilon}
f[r(\lambda),\theta(\lambda,\chi_0)]\,d\lambda\;.
\label{eq:torusave1D}
\end{equation}

Notice that in the resonant case, the average depends on the offset
phase $\chi_0$.  As such, if we imagine evolving from a non-resonant
to a resonant orbit, $\langle f\rangle$ will not change smoothly.
Instead, it will jump discontinuously as we move from the orbit in
which $\langle f\rangle$ does not depend on $\chi_0$ to the one where
it does so depend; and, the amount of jump will depend on the specific
value of $\chi_0$ we have chosen.

This discontinuous jumping behavior is an artifact of the infinite
time average, a limit which is of course irrelevant for a real
astrophysical inspiral.  A real system will spend some finite time
near any given orbit; if one wants to study averaged quantities, these
quantities should be averaged over something like that finite time.

As such, it should be understood that the infinite time averages that
we discuss in this paper are not intended to serve as tools to be used
for evolving extreme mass-ratio binaries through resonances.  For that
purpose, we instead advocate direct integration of the equations of
motion including self force --- without any averaging.  The infinite
time averaged rates of change we compute here are intended solely as
diagnostics of how a system's evolution is changed by resonant
physics, and how that change depends on the phase $\chi_0$.

\begin{widetext}

\subsection{Setup}
\label{sec:Qdotsetup}

We begin with the first line of Eq.\ (3.18) of S06.  It relates the
averaged rate of change of the Carter constant, per unit Mino time, to
the Kerr metric's Killing tensor $K^{\alpha\beta}$ and to a radiative
field $\Psi_{\rm rad}$ which is constructed from the perturbation to
the Kerr spacetime metric:
\begin{equation}
\left\langle\frac{dK}{d\lambda}\right\rangle \equiv
 \lim_{L \to
  \infty}\frac{1}{2L}\int_{-L}^L d\lambda \frac{dK}{d\lambda} =
 \lim_{L \to
  \infty}\frac{1}{2L}\int_{-L}^L d\lambda \left[2\Sigma
  K^{\alpha\beta}{\tilde u}_\alpha\partial_\beta \left(\frac{\Psi_{\rm
      rad}}{\Sigma}\right)\right]\biggl|_{x \to z(\lambda)}\;.
\label{eq:dKdlambda1}
\end{equation}
We refer the reader to S06 for a detailed derivation of
Eq.\ (\ref{eq:dKdlambda1}), and defer discussion of the radiative
field $\Psi_{\rm rad}(x)$ to Secs.\ {\ref{sec:nonresQdot}} and
{\ref{sec:resQdot}}.  The coordinate $x$ represents a general
spacetime field point; $x \to z(\lambda)$ means to take this general
point to the orbit's worldline $z(\lambda)$.

The other quantities appearing in Eq.\ (\ref{eq:dKdlambda1}) are as
follows: First, $K$ is a variant of the Carter constant, given by
\begin{equation}
K = Q + (L_z - a E)^2\;.
\label{eq:Kdef}
\end{equation}
It is related to the Kerr metric's Killing tensor by
\begin{equation}
K = K^{\alpha\beta}u_\alpha u_\beta\;,
\end{equation}
where
\begin{equation}
K^{\alpha\beta} = 2\Sigma m^{(\alpha} \bar m^{\beta)} -
a^2\cos^2\theta g^{\alpha\beta}\;.
\label{eq:Killingtensor}
\end{equation}
The tensor $g^{\alpha\beta}$ is the Kerr metric, and $m^\alpha$ are
components of the Newman-Penrose tetrad leg,
\begin{equation}
m^t = \frac{i a \sin\theta}{\sqrt{2}(r + ia\cos\theta)}\;,\quad m^r =
0\;, \quad m^\theta = \frac{1}{\sqrt{2}(r + ia\cos\theta)}\;,\quad
m^\phi = \frac{i\csc\theta}{\sqrt{2}(r + ia\cos\theta)}\;.
\label{eq:mcomponents}
\end{equation}
Overbar denotes complex conjugate.  The quantity $\tilde u_\alpha$ is
the 4-velocity promoted to a spacetime field:
\begin{equation}
(\tilde u_t, \tilde u_r, \tilde u_\theta, \tilde u_\phi) = (-E,
  \pm\sqrt{R(r)}/\Delta, \pm\sqrt{\Theta(\theta)}, L_z)\;,
\end{equation}
where $R(r)$ is defined in Eq.\ (\ref{eq:rdot}), and $\Theta(\theta)$
in Eq.\ (\ref{eq:thetadot}).  Notice that our $\tilde u_\theta$
differs from that used in S06.  This is due to a difference in the
definition of the potential $\Theta$ (it describes motion in $\theta$
here, but motion in $\cos\theta$ in S06).  The field $\tilde u_\alpha$
reduces exactly to the 4-velocity $u_\alpha$ when we take the limit of
the field point $x$ to the worldline $z(\lambda)$.

\subsection{General simplification}
\label{eq:Qdotsimplify}

We now take the first steps in simplifying Eq.\ (\ref{eq:dKdlambda1}).
These steps are the same for both resonant and non-resonant cases; we
specialize to those cases in Secs.\ {\ref{sec:nonresQdot}} and
{\ref{sec:resQdot}}.

We begin by focusing on the integrand of Eq.\ (\ref{eq:dKdlambda1}):
\begin{equation}
\left[2\Sigma K^{\alpha\beta}\tilde u_\alpha
  \partial_\beta\left(\frac{\Psi_{\rm rad}}{\Sigma}\right)\right]
\biggl|_{x \to z(\lambda)} = \left[4\Sigma^2m^{(\alpha}\bar
  m^{\beta)}\tilde u_\alpha \partial_\beta\left(\frac{\Psi^{\rm
      rad}}{\Sigma}\right) - 2\Sigma a^2\cos^2\theta \tilde u^\alpha
  \partial_\alpha\left(\frac{\Psi^{\rm
      rad}}{\Sigma}\right)\right]\biggl|_{x \to z(\lambda)}\;.
\label{eq:integrand1}
\end{equation}
Use the fact that $\tilde u^\alpha = u^\alpha$ in the limit $x \to
z(\lambda)$, and that $\Sigma u^\alpha = dx^\alpha/d\lambda$.
Expanding $m^{(\alpha}\bar m^{\beta)}$, we find
\begin{equation}
2\Sigma K^{\alpha\beta}\tilde u_\alpha
  \partial_\beta\left(\frac{\Psi_{\rm rad}}{\Sigma}\right) =
2\Sigma\left[\left(L_z - a\sin^2\theta E\right)
    \left(\csc^2\theta\partial_\phi + a\partial_t\right) +
    \frac{d\theta}{d\lambda}\partial_\theta\right]\left(\frac{\Psi^{\rm
      rad}}{\Sigma}\right) - 2 a^2\cos^2\theta
  \frac{d}{d\lambda}\left(\frac{\Psi^{\rm
      rad}}{\Sigma}\right)\;.
\label{eq:integrand2}
\end{equation}
[For brevity, we omit $x \to z(\lambda)$ in
  Eqs.\ (\ref{eq:integrand2}) and (\ref{eq:integrandcleanup}), though
  it should be understood that this limit is taken.]  The right-hand
side of Eq.\ (\ref{eq:integrand2}) can be simplified significantly by
combining the term in $d\theta/d\lambda$ with the final term:
\begin{eqnarray}
2\Sigma\frac{d\theta}{d\lambda}\partial_\theta\left(\frac{\Psi^{\rm
    rad}}{\Sigma}\right) - 2
a^2\cos^2\theta\frac{d}{d\lambda}\left(\frac{\Psi^{\rm
    rad}}{\Sigma}\right) &=&
2\frac{d\theta}{d\lambda}\partial_\theta\Psi_{\rm rad} - 2
\frac{\Psi^{\rm
    rad}}{\Sigma}\frac{d\theta}{d\lambda}\partial_\theta\Sigma
\nonumber\\
& & - 2a^2\frac{d}{d\lambda}\left(\cos^2\theta\frac{\Psi^{\rm
    rad}}{\Sigma}\right) + 2a^2\frac{\Psi^{\rm
    rad}}{\Sigma}\frac{d}{d\lambda}\left(\cos^2\theta\right)\;.
\label{eq:integrandcleanup}
\end{eqnarray}
The third term on the right-hand side of
Eq.\ (\ref{eq:integrandcleanup}) is a total derivative in
$d/d\lambda$.  Thanks to the periodic nature of all the relevant
terms, it will not contribute to an averaging integral of the form
(\ref{eq:dKdlambda1}), and may be discarded.  Using
\begin{equation}
\partial_\theta \Sigma = a^2\partial_\theta \cos^2\theta\;,
\quad\frac{d}{d\lambda}\cos^2\theta =
\frac{d\theta}{d\lambda}\partial_\theta\cos^2\theta\;,
\end{equation}
we see that the second and fourth terms on the right-hand side of
(\ref{eq:integrandcleanup}) cancel; only the term in
$\partial_\theta\Psi_{\rm rad}$ remains.  The integrand simplifies to
\begin{equation}
\left[2\Sigma K^{\alpha\beta}\tilde u_\alpha
  \partial_\beta\left(\frac{\Psi_{\rm
      rad}}{\Sigma}\right)\right]\biggl|_{x \to z(\lambda)} =
\left\{2\left[\left(L_z - a\sin^2\theta E\right)
  \left(\csc^2\theta\partial_\phi + a\partial_t\right) +
  \frac{d\theta}{d\lambda}\partial_\theta\right]\Psi_{\rm rad}
\right\}\biggl|_{x \to z(\lambda)}\;.
\label{eq:integrand3}
\end{equation}
The radiative field $\Psi_{\rm rad}$ can be broken into an ``out'' and
a ``down'' component:
\begin{equation}
\Psi_{\rm rad} = \Psi_{\rm rad}^{\rm out} + \Psi_{\rm rad}^{\rm down}\;.
\end{equation}
These two fields are in turn computed from mode functions
$\Phi_{lm\omega}$ (discussed in more detail momentarily) as follows:
\begin{equation}
\Psi_{\rm rad}^{\rm out}(x) = \int d\omega
\sum_{lm}\frac{1}{4i\omega^3} \left[\Phi_{lm\omega}^{\rm out}(x) \int
  d\lambda'\bar\Phi^{\rm out}_{lm\omega}[z(\lambda')]\right] +
\mbox{c.c.}\;,
\label{eq:Psi_out_rad}
\end{equation}
\begin{equation}
\Psi_{\rm rad}^{\rm down} = \int d\omega \sum_{lm}\frac{1}{4i\omega^2
  p_m} \left[\Phi_{lm\omega}^{\rm down}(x) \int d\lambda'\bar\Phi^{\rm
    down}_{lm\omega}[z(\lambda')]\right] + \mbox{c.c.}
\label{eq:Psi_down_rad}
\end{equation}
In Eq.\ (\ref{eq:Psi_down_rad}), $p_m = \omega - m\Omega_H$, where
$\Omega_H = a/2Mr_+$ is the angular velocity associated with the event
horizon.  The abbreviation ``c.c.'' means complex conjugate.  See S06
for further discussion and derivation of these forms of the fields
$\Psi^{\rm out}_{\rm rad}$ and $\Psi^{\rm down}_{\rm rad}$.  We will
largely focus on the ``out'' field, which is related to radiation at
${\cal I}^+$.  Extension to the ``down'' field, related to radiation
on the event horizon, is straightforward.

To proceed, we use two equivalent forms for
$\Phi^{\rm out}_{lm\omega}(x)$ evaluated in the limit $x \to
z(\lambda)$; both are described in more detail in S06.
%The first is a
%simple Fourier series expansion:
%\begin{equation}
%\Phi^{\rm out}_{lm\omega}[z(\lambda)] =
%\frac{\Gamma}{2\pi}\sum_{nk}\bar Z^H_{lmkn} e^{i\Gamma(\omega -
%  \omega_{mkn})\lambda}\;.
%\label{eq:modefunc_fourier}
%\end{equation}
%This is Eq.\ (3.11) of S06, translated into our notation; $\Gamma$ is
%the factor introduced in Sec.\ {\ref{sec:resonances}} that converts
%the mean accumulation of Mino time to the mean accumulation of
%coordinate time. \note{Eanna, I did not agree with your sign flip
%  here.  Details in my email to you.}
The first is up to a constant factor the complex conjugate of the
integrand in the expression
(\ref{eq:Z2}) for $Z^H_{lm\omega}$:
\begin{equation}
\Phi^{\rm out}_{lm\omega}[z(\lambda)] = {\bar J}^H_{lm\omega}(\lambda)
e^{-i\lambda(\Gamma \omega -
  m \Upsilon_\phi)}\;.
\label{eq:modefunc_fourier}
\end{equation}
Here $\Gamma$ is the factor introduced in Sec.\ {\ref{sec:resonances}}
that converts the mean accumulation of Mino time to the mean
accumulation of coordinate time.  Equation (\ref{eq:modefunc_fourier})
is Eq.\ (3.11) of S06, translated into our notation\footnote{Note that
  there are two errors in Eq.\ (3.11) of S06: the sign of the
  exponential is flipped, and the coefficients $Z$ are of the usual
  type (\protect{\ref{eq:Zcoefa}}) rather than the required more
  general type (\protect{\ref{eq:Zcoef}}) with $\omega \ne
  \omega_{mkn}$.  See Eq.\ (\protect{\ref{eq:Phioutfull}}) below.};
the scalar-case version of this equation is Eq.\ (9.20) of
Ref.\ \cite{dfh05}.  Using the Fourier series expansion
(\ref{eq:Jdecomp}) of $J^H_{lm\omega}$, integrating with respect to
$\lambda$, and combining with the definitions (\ref{eq:Zcoef1}) and
(\ref{eq:Zcoefa}) gives
\begin{equation}
\int d\lambda'\bar\Phi^{\rm out}_{lm\omega}[z(\lambda')] =
\sum_{nk}Z^H_{lmkn}\delta(\omega - \omega_{mkn})\;,
\end{equation}
and so
\begin{equation}
\Psi_{\rm rad}^{\rm out}(x) = \int d\omega \left[\,
  \sum_{lmkn}\frac{Z^H_{lmkn}\delta(\omega -
    \omega_{mkn})}{4i\omega^3} \Phi_{lm\omega}^{\rm out}(x)\right] +
\mbox{c.c.}
\label{eq:Psi_out_rad2}
\end{equation}
A similar simplification describes $\Psi_{\rm rad}^{\rm down}(x)$.
Combining this with Eqs.\ (\ref{eq:dKdlambda1}) and
(\ref{eq:integrand3}), we obtain
\begin{eqnarray}
\left\langle\frac{dK^\infty}{d\lambda}\right\rangle &=&
\left\langle
\sum_{lmkn}\frac{Z^H_{lmkn}}{2i\omega_{mkn}^3}
%\right.
%\nonumber\\ & &\qquad\qquad
%\left. \times
\left\{\left[(\csc^2\theta L_z - aE)\partial_\phi + a(L_z -
  aE\sin^2\theta) \partial_t +
  \frac{d\theta}{d\lambda}\partial_\theta\right] \Phi^{\rm
  out}_{lmkn}\right\} + \mbox{c.c.} \right\rangle,
\label{eq:dKdlambda2}
\end{eqnarray}
where $\Phi^{\rm out}_{lmkn} \equiv \Phi^{\rm out}_{lm\omega_{mkn}}$.
(The superscript ``$\infty$'' is because we focus on the ``out''
field.)

We next manipulate the term in $\partial_\theta$ in Eq.\
(\ref{eq:dKdlambda2}), by invoking the second form for $\Phi^{\rm
  out}_{lm\omega}(x)$, which is
\begin{equation}
\Phi^{\rm out}_{lmkn}(x) = f_{lmkn}(r,\theta) e^{i m \phi} e^{-i
  \omega_{mkn} t}\;.
\end{equation}
The value of $f_{lmkn}(r,\theta)$ is not important for our purposes;
see S06 [Eq.\ (3.20) and nearby text] for further details.  We have
changed notation from S06 slightly to highlight the fact that this
function depends on $l$, $m$, $k$, and $n$; this is important for
generalizing to resonant orbits.  We now evaluate on the worldline $x
\to z(\lambda)$, and use the following explicit representations of the
motions in $t$ and $\phi$:
\begin{eqnarray}
t(\lambda) &=& \Gamma \lambda + \Delta t_r(\lambda) + \Delta
t_\phi(\lambda), \nonumber \\
\phi(\lambda) &=& \Upsilon_\phi \lambda + \Delta \phi_r(\lambda) + \Delta
\phi_\phi(\lambda),
\end{eqnarray}
cf.\ Eqs.\ (\ref{tdecompos}) and (\ref{phidecompos}) above and Sec.\ 3
of Ref.\ \cite{dfh05}.  Here the function $\Delta t_r$ is periodic
with period $\Lambda_r$ and $\Delta t_\theta$ is periodic with period
$\Lambda_\theta$, etc.  This gives
\begin{equation}
\Phi^{\rm out}_{lmkn}(\lambda) = f_{lmkn}[r(\lambda),\theta(\lambda)]
\exp \left\{- i \lambda( k \Upsilon_\theta  + n \Upsilon_r)
     - i \omega_{mkn} [\Delta t_r(\lambda) + \Delta t_\theta(\lambda)]
     + i m [\Delta \phi_r(\lambda) + \Delta \phi_\theta(\lambda)]
 \right\}.
\label{eq:phiout1}
\end{equation}
We next define a mode function of two variables $(\lambda^r,
\lambda^\theta)$ by
\begin{equation}
\Phi^{\rm out}_{lmkn}(\lambda^r,\lambda^\theta) \!=\!
f_{lmkn}[r(\lambda^r),\theta(\lambda^\theta)]
\exp \left\{- i k \Upsilon_\theta \lambda^\theta -i n \Upsilon_r \lambda^r
     - i \omega_{mkn} [\Delta t_r(\lambda^r) + \Delta t_\theta(\lambda^\theta)]
     + i m [\Delta \phi_r(\lambda^r) + \Delta \phi_\theta(\lambda^\theta)]
 \right\}.
\label{Phiexplicit}
\end{equation}
This function is determined uniquely by the following three
properties: First, it reduces to the expression (\ref{eq:phiout1})
when evaluated at $\lambda^r = \lambda^\theta = \lambda$; second, it
is biperiodic, with a period of $\Lambda^r$ in $\lambda^r$, and of
$\Lambda^\theta$ in $\lambda^\theta$; and third, it is a continuous
function of the geodesic's parameters.  The first two properties are
sufficient to guarantee uniqueness for non-resonant orbits, but not
for resonant orbits since the different periodicities become
degenerate.  Adding the third property is sufficient to restore
uniqueness for all orbits, since resonant orbits form a set of measure
zero in the phase space.  See Refs.\ \cite{dh04,dfh05} for more
details on the mapping between functions of $\lambda$ and functions of
$(\lambda^r,\lambda^\theta)$.

Next, differentiating the explicit expression (\ref{Phiexplicit}) with
respect to $\lambda^\theta$, we obtain the following identity relating
the differential operator $d/d\lambda^\theta$ and the partial
derivative operators $\partial_\theta$, $\partial_t$ and
$\partial_\phi$ acting on $\Phi^{\rm out}_{lmkn}$:
\begin{equation}
\frac{d\theta}{d\lambda} \partial_\theta =
\frac{d}{d\lambda^\theta} + i k \Upsilon_\theta
-\frac{d\Delta t_\theta}{d\lambda^\theta}\partial_t -
\frac{d \Delta \phi_\theta}{d\lambda^\theta}\partial_\phi.
\label{eq:ddlambdathetatot}
\end{equation}
We now use the identity (\ref{eq:ddlambdathetatot})
to substitute for the $(d\theta/d\lambda)\partial_\theta$
term in Eq.\ (\ref{eq:dKdlambda2}).  This yields
\begin{eqnarray}
& &\left\langle\frac{dK^\infty}{d\lambda}\right\rangle =\left\langle
  \sum_{lmkn}\frac{Z^H_{lmkn}}{2i\omega_{mkn}^3}
\right.
\nonumber\\ & & \qquad\quad
\left.
  \times \left\{\left[\left(\csc^2\theta L_z - aE -
    \frac{d\Delta \phi_\theta}{d\lambda^\theta}\right)\partial_\phi +
  \left(a L_z -
    a^2 E\sin^2\theta - \frac{d\Delta
      t_\theta}{d\lambda^\theta}\right) \partial_t + i k
  \Upsilon_\theta + \frac{d}{d\lambda^\theta} \right] \Phi^{\rm out}_{lmkn}\right\} + \mbox{c.c.} \right\rangle.
  \nonumber\\
\label{eq:dKdlambda3}
\end{eqnarray}
Using Eqs.\ (3.43) and (3.58) of Ref.\ \cite{dfh05}
it is not difficult to show that
\begin{eqnarray}
& &\csc^2\theta L_z - a E - \frac{d\Delta \phi_\theta}{d\lambda^\theta} =
\langle\csc^2\theta L_z - a E\rangle =
\langle\csc^2\theta\rangle L_z - a E\;,
\label{eq:ave1}\\
& &a L_z - a^2 E\sin^2\theta - \frac{d\Delta t_\theta}{d\lambda^\theta} =
\langle a L_z - a^2 E\sin^2\theta\rangle =
a L_z - a^2 E\langle\sin^2\theta\rangle\;.
\label{eq:ave2}
\end{eqnarray}
Combining this with Eq.\ (\ref{eq:dKdlambda3}) and using the
replacements $\partial_\phi \to i m$, $\partial_t \to - i\omega_{mkn}$
gives
\begin{equation}
\left\langle\frac{dK^\infty}{d\lambda}\right\rangle =\left\langle
  \sum_{lmkn}\frac{Z^H_{lmkn}}{2i\omega_{mkn}^3}
  \left\{\left[i {\cal M}_{mkn} + i k
  \Upsilon_\theta + \frac{d}{d\lambda^\theta} \right] \Phi^{\rm out}_{lmkn}\right\} + \mbox{c.c.} \right\rangle,
\label{eq:dKdlambda3a}
\end{equation}
where we have defined
\begin{equation}
{\cal M}_{mkn}  = m (\langle\csc^2\theta\rangle L_z - a E) - a
\omega_{mkn} ( L_z - a E\langle\sin^2\theta\rangle).
\label{eq:calMdef}
\end{equation}
Next, from Eqs.\ (\ref{eq:modefunc_fourier}), (\ref{eq:Jdecomp}), (\ref{eq:Zcoef1}) and
(\ref{eq:Zcoefa}) we obtain an expression
for $\Phi^{\rm out}_{lmkn}(\lambda)$.  Extending this to a function of
$\lambda^r,\lambda^\theta$ as above gives
\begin{equation}
\Phi^{\rm out}_{lmkn}(\lambda^r,\lambda^\theta) = \frac{\Gamma}{2 \pi}
\sum_{\Delta n,\Delta k} {\bar Z}^H_{\omega_{mkn} lm k+\Delta k, n+\Delta n} e^{
i \Delta k \Upsilon_\theta \lambda^\theta} e^{ i \Delta n \Upsilon_r
\lambda^r}.
\label{eq:Phioutfull}
\end{equation}
Combining this with Eq.\ (\ref{eq:dKdlambda3a}) yields the final
result
\begin{equation}
\left\langle\frac{dK^\infty}{d\lambda}\right\rangle =\left\langle
\frac{\Gamma}{4 \pi}  \sum_{lmkn} \sum_{\Delta k,\Delta n}
  \left[{\cal M}_{mkn} +  k
  \Upsilon_\theta + \Delta k \Upsilon_\theta \right]
 \frac{Z^H_{lmkn}}{\omega_{mkn}^3}
{\bar Z}^H_{\omega_{mkn} lm k+\Delta k, n+\Delta n}
e^{i \Delta k \Upsilon_\theta \lambda^\theta} e^{ i \Delta n \Upsilon_r \lambda^r}+ \mbox{c.c.} \right\rangle.
\label{eq:dKdlambda3b}
\end{equation}
Here it is understood that the averaging procedure is to first
evaluate at $\lambda^r = \lambda^\theta \equiv \lambda$ and then
average over $\lambda$.  In Sec.\ {\ref{sec:nonresQdot}}, we evaluate
this average for non-resonant orbits, and reproduce the results of
S06.  In Sec.\ {\ref{sec:resQdot}}, we do so for a resonant orbit and
find an appropriately modified variant of their formula.

\subsection{Non-resonant result}
\label{sec:nonresQdot}

We evaluate the expression (\ref{eq:dKdlambda3b}) at $\lambda^r =
\lambda^\theta \equiv \lambda$ and then evaluate the average over
$\lambda$ defined by Eq.\ (\ref{eq:dKdlambda1}).  The term labeled by
$\Delta n$, $\Delta k$ is proportional to
\begin{equation}
 \lim_{L \to
  \infty}\frac{1}{2L}\int_{-L}^L d\lambda\, e^{i \Delta k
  \Upsilon_\theta \lambda} e^{ i \Delta n \Upsilon_r \lambda} =  \lim_{L \to
  \infty} {\rm Si}[ (\Delta k \Upsilon_\theta + \Delta n \Upsilon_r) L],
\end{equation}
where ${\rm Si}(x) = \sin(x)/x$.  Since the frequencies
$\Upsilon_\theta$ and $\Upsilon_r$ are incommensurate for non-resonant
orbits, the combination $\Delta k \Upsilon_\theta + \Delta n
\Upsilon_r$ will be nonvanishing for $(\Delta k,\Delta n) \ne (0,0)$,
and the right hand side will vanish.  Thus the only non-vanishing term
will be the term with $\Delta n = \Delta k = 0$.
Another way to think about this is that we are averaging over a curve
which is ergodically filling up the torus parameterized by $\lambda^r$
and $\lambda^\theta$, and so the curve average can be replaced by an
average over the torus,
\begin{equation}
\lim_{L\to \infty}\frac{1}{2L}\int_{-L}^L
\ldots
d\lambda \to
\frac{\Upsilon_\theta\Upsilon_r}{(2\pi)^2}\int_0^{2\pi/\Upsilon_\theta}
\int_0^{2\pi/\Upsilon_\theta}
\ldots
\,d\lambda^rd\lambda^\theta\;.
\label{eq:averaging}
\end{equation}
Applying this torus average to the expression (\ref{eq:dKdlambda3b})
again forces $\Delta n = \Delta k =0$.  Now using the definition
(\ref{eq:Zcoefa}) we obtain the final result
\begin{equation}
\left\langle \frac{dK^\infty}{d\lambda}\right\rangle = \Gamma
\sum_{lmkn}\frac{|\check{Z}^H_{lmkn}|^2}{4\pi\omega_{mkn}^3}
\left[{\cal M}_{mkn} + k\Upsilon_\theta \right] +
\mbox{c.c.}
\label{eq:dKdlambda4}
\end{equation}
Because all the terms on the right-hand side of (\ref{eq:dKdlambda4})
are real, the complex conjugate simplifies to an overall factor of
two.  We take the long-time average, so
\begin{equation}
\left\langle\frac{dK}{d\lambda}\right\rangle = \Gamma
\left\langle\frac{dK}{dt}\right\rangle\;.
\label{eq:Kdot1}
\end{equation}
Further, by Eq.\ (\ref{eq:Kdef}),
\begin{equation}
\frac{dK}{dt} = \frac{dQ}{dt} + 2\left(a E -
L_z\right)\left(a\frac{dE}{dt} - \frac{dL_z}{dt}\right)\;.
\label{eq:Kdot2}
\end{equation}
Combining Eqs.\
(\ref{eq:EdotInfnonres}), (\ref{eq:LzdotInfnonres}),
(\ref{eq:calLdef}), (\ref{eq:calMdef}) together with Eqs.\
(\ref{eq:dKdlambda4}), (\ref{eq:Kdot1}), and
(\ref{eq:Kdot2}), we finally obtain
\begin{eqnarray}
\left\langle\frac{dQ^\infty}{dt}\right\rangle &=&
\sum_{lmkn}\frac{|\check{Z}_{lmkn}^H|^2}{2\pi\omega_{mkn}^3} \left(m
    \langle\cot^2\theta\rangle L_z - a^2\omega_{mkn}
    \langle\cos^2\theta\rangle E + k\Upsilon_\theta\right)
\nonumber\\
&\equiv& 2\sum_{lmkn} \frac{\dot
  E_{lmkn}^\infty}{\omega_{mkn}}\left({\cal L}_{mkn} + k\Upsilon_\theta\right)
\;.
\label{eq:Qdotnonres_Inf}
\end{eqnarray}
The quantity ${\cal L}_{mkn}$ is defined in Eq.\ (\ref{eq:calLdef}).
A similar calculation focusing on the ``down'' modes yields
\begin{eqnarray}
\left\langle\frac{dQ^H}{dt}\right\rangle &=&
\sum_{lmkn}\frac{\alpha_{lmkn}
  |\check{Z}_{lmkn}^\infty|^2}{2\pi\omega_{mkn}^3} \left(m
\langle\cot^2\theta\rangle L_z - a^2\omega_{mkn}
\langle\cos^2\theta\rangle E + k\Upsilon_\theta\right) \nonumber\\ &=&
2\sum_{lmkn}\frac{{\dot E}^H_{lmkn}}{\omega_{mkn}}
\left({\cal L}_{mkn} + k\Upsilon_\theta\right)\;.
\label{eq:Qdotnonres_H}
\end{eqnarray}
The factor $\alpha_{lmkn}$ is introduced in Sec.\ {\ref{sec:fluxes}};
on the second line, we have used Eqs.\ (\ref{eq:EdotHnonres}) and
(\ref{eq:calLdef}).  Equations (\ref{eq:Qdotnonres_Inf}) and
(\ref{eq:Qdotnonres_H}) are the same (modulo minor changes in
notation) as Eq.\ (3.26) of S06.

\subsection{Resonant $\dot Q$}
\label{sec:resQdot}

We now return to the general formula (\ref{eq:dKdlambda3b}) evaluated
at $\lambda^r = \lambda^\theta = \lambda$ and compute the average over
$\lambda$ for
the case of resonant orbits.  Before evaluating this average
we first simplify the sums over $\Delta k$
and $\Delta n$ by rewriting them in terms of $k' = k + \Delta k$, $n'
= n + \Delta n$.  We also make the replacements
\begin{equation}
\sum_{kn} \to \sum_N \ \sum_{(k,n)_N}, \ \ \ \ \ \ \sum_{k'n'} \to
\sum_{N'} \ \sum_{(k',n')_{N'}},
\end{equation}
where the indicated sums are taken over $k,n$ satisfying $k
\beta_\theta + n \beta_r = N$ and over $k',n'$ satisfying $k'
\beta_\theta + n' \beta_r = N'$.
We note that the quantities ${\cal M}_{mkn}$ and $\omega_{mkn}$
depend on $k$ and $n$ only through $N$, and write these
as ${\cal M}_{mN}$ and $\omega_{mN}$.  Finally using the definition
(\ref{eq:Zres_expand}) of the amplitudes ${\cal Z}^\star_{lmN}$, the
expression (\ref{eq:dKdlambda3b}) reduces to
\begin{equation}
\left\langle\frac{dK^\infty}{d\lambda}\right\rangle =\left\langle
\frac{\Gamma}{4 \pi}  \sum_{lmN} \sum_{N'} \sum_{(k',n')_{N'}}
  \left[{\cal M}_{mN} +  k'
  \Upsilon_\theta \right]
 \frac{{\cal Z}^H_{lmN}}{\omega_{mN}^3}
{\bar Z}^H_{\omega_{mkn}lm k' n'}
e^{i \Delta k \Upsilon_\theta \lambda} e^{ i \Delta n \Upsilon_r \lambda}+ \mbox{c.c.} \right\rangle.
\label{eq:dKdlambda3c}
\end{equation}
Next we note that the argument of the exponential is
\begin{equation}
i \lambda( \Delta k \Upsilon_\theta + \Delta n \Upsilon_r) =
i \lambda \Upsilon (\Delta k \beta_\theta + \Delta n \beta_r) = i
\lambda \Upsilon (N'-N).
\end{equation}
Evaluating the average over $\lambda$ enforces $N = N'$, and the
result is
\begin{equation}
\left\langle\frac{dK^\infty}{d\lambda}\right\rangle =
\frac{\Gamma}{4 \pi}  \sum_{lmN} \sum_{(k',n')_{N}}
  \left[{\cal M}_{mN} +  k'
  \Upsilon_\theta \right]
 \frac{{\cal Z}^H_{lmN}}{\omega_{mN}^3}
{\bar Z}^H_{\omega_{mkn}lm k' n'}+ \mbox{c.c.} .
\label{eq:dKdlambda3d}
\end{equation}

Now since $\omega_{mkn} = \omega_{mN} = \omega_{mN'}$, the factor of
${\bar Z}^H_{\omega_{mkn}lm k' n'}$ can be simplified to
${\bar Z}^H_{lm k' n'}$.
The expression (\ref{eq:dKdlambda3d}) can then simplified further by
defining the
new amplitude
\begin{equation}
{\cal Y}^H_{lmN}(\chi_0) = \sum_{(k,n)_N}
k Z^H_{lmkn}(\chi_0) = \sum_{(k,n)_N} k e^{i\xi_{mkn}(\chi_0)} \check{Z}^H_{lmkn}\;.
\label{eq:Ydef}
\end{equation}
Compare this with Eq.\ (\ref{eq:Zres_expand}): ${\cal
  Y}^H_{lmN}(\chi_0)$ is similar to ${\cal Z}_{lmN}(\chi_0)$, but with
each $Z^H_{lmkn}$ weighted by $k$.  In terms of this new amplitude the
result simplifies to
\begin{eqnarray}
\left\langle \frac{dK^\infty}{d\lambda}\right\rangle &=&
\sum_{lmN}\frac{\Gamma}{4\pi\omega_{mN}^3}\left[ {\cal M}_{mN} |{\cal
    Z}^H_{lmN}(\chi_0)|^2
%\right.
%\nonumber\\
%& &\qquad\qquad\qquad\qquad\qquad\qquad\qquad
% \left.
+\Upsilon_\theta
{\cal Z}^H_{lmN}(\chi_0)\bar{\cal Y}^H_{lmN}(\chi_0)\right] +
\mbox{c.c.}
\label{eq:dKdlambda_res3}
\end{eqnarray}
Applying Eqs.\ (\ref{eq:Kdot1}) and (\ref{eq:Kdot2}), we at last
find the rate of change of $Q$ for a resonant orbit:
\begin{equation}
\left\langle\frac{dQ^\infty}{dt}\right\rangle =
\sum_{lmN}\frac{1}{2\pi\omega_{mN}^3}\left\{ {\cal L}_{mN} |{\cal
    Z}^H_{lmN}(\chi_0)|^2+ \Upsilon_\theta \mbox{Re}\left[{\cal Z}^H_{lmN}(\chi_0)
\bar{\cal Y}^H_{lmN}(\chi_0)\right]\right\}\;,
\label{eq:Qdotres_Inf1}
\end{equation}
where ${\cal L}_{mN}$ is the same as ${\cal L}_{mkn}$, but with
$\omega_{mkn} \to \omega_{mN}$.  Repeating this exercise for the
``down'' modes yields
\begin{equation}
\left\langle\frac{dQ^H}{dt}\right\rangle =
\sum_{lmN}\frac{\alpha_{lmN}}{2\pi\omega_{mN}^3}\left\{{\cal L}_{mN} |{\cal
  Z}^\infty(\chi_0)|^2 + \Upsilon_\theta
\mbox{Re}\left[{\cal Z}^\infty_{lmN}(\chi_0) \bar{\cal
    Y}^\infty_{lmN}(\chi_0)\right]\right\}\;.
\label{eq:Qdotres_H1}
\end{equation}
\end{widetext}

It is interesting to compare our final result for the on-resonance
evolution of $Q$, Eqs.\ (\ref{eq:Qdotres_Inf1}) and
(\ref{eq:Qdotres_H1}), with the equivalent results for the
non-resonant case, Eqs.\ (\ref{eq:Qdotnonres_Inf}) and
(\ref{eq:Qdotnonres_H}).  The first two terms in both expressions for
$\langle dQ/dt\rangle$ are essentially the same; going from the
non-resonant case to the resonant case is simply a matter of promoting
the 4-index non-resonant amplitude $Z^\star_{lmkn}$ to the 3-index
resonant amplitude ${\cal Z}^\star_{lmN}$.

The final term in the two cases is quite different, however.  In the
non-resonant case, the final term is proportional to
$k\Upsilon_\theta$.  In the resonant case, the index $k$ cannot appear
in the final result, which can only depend on the indices $l$, $m$,
and $N$.  This is accounted for in the definition of the amplitude
${\cal Y}^\star_{lmN}$, Eq.\ (\ref{eq:Ydef}).  In both the
non-resonant and the resonant cases, this final term arises from the
action of the operator $(d\theta/d\lambda)\partial_\theta$ on the
radiative field $\Psi_{\rm rad}$ [see Eq.\ (\ref{eq:integrand3})].

As Appendix \ref{app:equivalence} made clear, the 3-index amplitude
${\cal Z}^\star_{lmN}$ can be computed directly as a 1-D integral,
Eq.\ (\ref{eq:Zres_int}), or can be computed as a sum of 4-index
integrals, Eq.\ (\ref{eq:Zres_expand}), each of which is computed from
the 2-D integral (\ref{eq:Zcoef}).  Our definition (\ref{eq:Ydef}) of
${\cal Y}^\star_{lmN}$ is clearly analogous to
Eq.\ (\ref{eq:Zres_expand}), writing this 3-index amplitude as a sum
over 4-index amplitudes.

Might it be possible to compute the 3-index amplitude directly, in a
manner analogous to Eq.\ (\ref{eq:Zres_int})?  We believe the answer
is yes: We simply need to propagate the operator
$(d\theta/d\lambda)\partial_\theta$ under the integral sign in
Eq.\ (\ref{eq:Zres_int}).  In other words, we speculate that
\begin{equation}
{\cal Y}^\star_{lmN}(\chi_0) \speculate
\frac{\Upsilon}{\Gamma}\int_0^{2\pi/\Upsilon}\!\!\!d\lambda\,
\frac{d\theta}{d\lambda}\partial_\theta
J^\star_{lm\omega}[r(\lambda),\theta(\lambda,\chi_0)]e^{i N
  \Upsilon\lambda}\;.
\label{eq:Ycof_int}
\end{equation}
We have not yet tested this.

\section{Rate of change of $E$ and $L_z$ by dissipative self force}
\label{app:edotlzdot}

With $\langle dQ/dt\rangle$ due to the dissipative self force now
understood, it is a relatively simple matter to likewise compute
$\langle dE/dt\rangle$ and $\langle dL_z/dt\rangle$.  Our calculation
again closely follows S06; the only important changes are updates to
the notation that we use, and a careful analysis of resonances.  The
results we find are identical to the fluxes of energy and angular
momentum carried by gravitational waves, exactly as
Ref.\ {\cite{qw99}} leads us to expect.

\subsection{Setup}

Our starting point is Eq.\ (3.7) of S06, which in our notation becomes
\begin{eqnarray}
\left\langle\frac{dE}{d\lambda}\right\rangle &\equiv& \lim_{L
  \to \infty}\frac{1}{2L}\int_{-L}^L d\lambda \frac{dE}{d\lambda}
\nonumber\\
&=& -\lim_{L \to \infty}\frac{1}{2L}\int_{-L}^L d\lambda \left[\partial_t
  \Psi_{\rm rad}\right]\Bigl|_{x \to z(\lambda)}\;.
\end{eqnarray}
This equation is derived by averaging over long times the dissipative
self force contracted with the time Killing vector.  Terms
corresponding to total derivatives are discarded thanks to the
periodic nature of the underlying functions.  If we replace
$-\partial_t$ with $\partial_\phi$, we obtain $\langle
dL_z/d\lambda\rangle$.

As in Appendix {\ref{app:qdot}}, we'll focus on the ``out'' fields;
extension to ``down'' is straightforward.  Using
Eq.\ (\ref{eq:Psi_out_rad2}),
\begin{equation}
\left\langle\frac{dE^\infty}{d\lambda}\right\rangle =
-\left\langle\sum_{lmkn}\frac{Z^H_{lmkn}}{4i\omega_{mkn}^3}\partial_t\Phi^{\rm
  out}_{lmkn} + {\rm c.c.} \right\rangle\;.
\end{equation}
The harmonic behavior of the mode functions means that
$\partial_t\Phi^{\rm out}_{lmkn} = -i\omega_{mkn}\Phi^{\rm
  out}_{lmkn}$:
\begin{equation}
\left\langle\frac{dE^\infty}{d\lambda}\right\rangle =
\left\langle\sum_{lmkn}\frac{Z^H_{lmkn}}{4\omega_{mkn}^2}\Phi^{\rm
  out}_{lmkn} + {\rm c.c.} \right\rangle
\end{equation}
Using Eq.\ (\ref{eq:Phioutfull}), this becomes
\begin{eqnarray}
& &\left\langle\frac{dE^\infty}{d\lambda}\right\rangle =
\Biggl\langle\frac{\Gamma}{8\pi}\sum_{lmkn}\sum_{\Delta k,\Delta n}
\frac{Z^H_{lmkn}}{\omega_{mkn}^2}
\times
\nonumber\\
& &
 \bar Z^H_{\omega_{mkn}lmk+\Delta k,n+\Delta n}
e^{i\Delta k\Upsilon_\theta\lambda^\theta}
e^{i\Delta n\Upsilon_r\lambda^r} + {\rm c.c.}\Biggr\rangle\;.
\nonumber\\
\label{eq:dEdlambda}
\end{eqnarray}
Likewise, using $\partial_\phi\Phi^{\rm out}_{lmkn} = im\Phi^{\rm
  out}_{lmkn}$, we have
\begin{eqnarray}
& &\left\langle\frac{dL_z^\infty}{d\lambda}\right\rangle =
\Biggl\langle\frac{\Gamma}{8\pi}\sum_{lmkn}\sum_{\Delta k,\Delta n}
m\frac{Z^H_{lmkn}}{\omega_{mkn}^3}
\times
\nonumber\\
& &
 \bar Z^H_{\omega_{mkn}lmk+\Delta k,n+\Delta n}
e^{i\Delta k\Upsilon_\theta\lambda^\theta}
e^{i\Delta n\Upsilon_r\lambda^r} + {\rm c.c.}\Biggr\rangle\;.
\nonumber\\
\label{eq:dLzdlambda}
\end{eqnarray}
As in App.\ {\ref{app:qdot}}, the averaging procedure we use is to
evaluate at $\lambda^r = \lambda^\theta = \lambda$, and then to
average over $\lambda$.  We do this first for non-resonant and then
for resonant orbits.

\subsection{Non-resonant results}

As in Appendix {\ref{sec:nonresQdot}}, we use the fact that
\begin{eqnarray}
 \lim_{L \to
  \infty}\frac{1}{2L}\int_{-L}^L d\lambda\, e^{i \Delta k
  \Upsilon_\theta \lambda} e^{ i \Delta n \Upsilon_r \lambda} &=&
\nonumber\\
& &
\!\!\!\!\!\!\!\!\!\!\!\!\!\!\!\!\!\!\!\!\!\!\!\!
\!\!\!\!\!\!\!\!\!\!\!\!\!\!\!\!\!\!\!\!\!\!\!\!
\!\!\!\!
  \lim_{L \to \infty} {\rm Si}[ (\Delta k \Upsilon_\theta + \Delta n
    \Upsilon_r) L]\;,
\end{eqnarray}
where ${\rm Si}(x) = \sin(x)/x$.  For non-resonant orbits, the
incommensurability of $\Upsilon_\theta$ and $\Upsilon_r$ means that
the only nonvanishing term is $\Delta n = \Delta k = 0$, and we deduce
that
\begin{eqnarray}
\left\langle\frac{dE^\infty}{dt}\right\rangle &=&
\sum_{lmkn}
\frac{|Z^H_{lmkn}|^2}{4\pi\omega_{mkn}^2}\;,
\label{eq:EdotInfnonres2}
\\
\left\langle\frac{dL_z^\infty}{dt}\right\rangle &=&
\sum_{lmkn}
\frac{m|Z^H_{lmkn}|^2}{4\pi\omega_{mkn}^3}\;.
\label{eq:LzdotInfnonres2}
\end{eqnarray}
We have used the fact that the factor $\Gamma$ converts, on a
long-time average basis, derivatives in $\lambda$ to derivatives in
$t$.  Repeating this calculation for the ``down'' modes, we find
\begin{eqnarray}
\left\langle\frac{dE^H}{dt}\right\rangle &=&
\sum_{lmkn}\alpha_{lmkn}
\frac{|Z^\infty_{lmkn}|^2}{4\pi\omega_{mkn}^2}\;,
\label{eq:EdotHnonres2}
\\
\left\langle\frac{dL_z^H}{dt}\right\rangle &=&
\sum_{lmkn}\alpha_{lmkn}
\frac{m|Z^\infty_{lmkn}|^2}{4\pi\omega_{mkn}^3}\;.
\label{eq:LzdotHnonres2}
\end{eqnarray}
The factor $\alpha_{lmkn}$ is discussed in Sec.\ {\ref{sec:fluxes}}.
Equations (\ref{eq:EdotInfnonres2})--(\ref{eq:LzdotHnonres2}) are
identical to Eqs.\ (\ref{eq:EdotInfnonres})--(\ref{eq:LzdotHnonres}).

\subsection{Resonant results}

As in App.\ {\ref{sec:resQdot}}, we first modify the sums by rewriting
them in terms of $k' = k + \Delta k$, $n' = n + \Delta n$, and
make the replacements
\begin{equation}
\sum_{kn} \to \sum_N\sum_{(k,n)_N}\;,\qquad
\sum_{k'n'} \to \sum_{N'}\sum_{(k',n')_N'}\;,
\end{equation}
where the sums are taken over pairs satisfying $k\beta_\theta +
n\beta_r = N$ and $k'\beta_\theta + n'\beta_r = N'$.  We use the fact
that $\omega_{mkn}$ depends on $k$ and $n$ only through $N$ to replace
it with $\omega_{mN}$, and use the definition (\ref{eq:Zres_expand})
of ${\cal Z}^\star_{lmN}$ to write (\ref{eq:dEdlambda}) as
\begin{eqnarray}
& &\left\langle\frac{dE^\infty}{d\lambda}\right\rangle =
\Biggl\langle\frac{\Gamma}{8\pi}\sum_{lmN}\sum_{N'}\sum_{(k',n')_{N'}}
\frac{{\cal Z}^H_{lmN}}{\omega_{mN}^2}
\times
\nonumber\\
& &
 \bar Z^H_{\omega_{mkn}lmk'n'}e^{i\Delta k\Upsilon_\theta\lambda}
e^{i\Delta n\Upsilon_r\lambda} + {\rm c.c.}\Biggr\rangle\;.
\nonumber\\
\label{eq:dEdlambdares}
\end{eqnarray}
A similar expression describes $\langle dL^\infty_z/dt\rangle$.  Using
the same logic as follows Eq.\ (\ref{eq:dKdlambda3c}), we see that
averaging over $\lambda$ enforces $N = N'$, and we obtain
\begin{eqnarray}
\left\langle\frac{dE^\infty}{dt}\right\rangle &=&
\sum_{lmN}
\frac{|{\cal Z}^H_{lmN}|^2}{4\pi\omega_{mN}^2}\;,
\label{eq:EdotInfres2}
\\
\left\langle\frac{dL_z^\infty}{dt}\right\rangle &=&
\sum_{lmN}
\frac{m|{\cal Z}^H_{lmN}|^2}{4\pi\omega_{mN}^3}\;.
\label{eq:LzdotInfres2}
\end{eqnarray}
The same analysis for the ``down'' modes yields
\begin{eqnarray}
\left\langle\frac{dE^H}{dt}\right\rangle &=&
\sum_{lmN}\alpha_{lmN}
\frac{|{\cal Z}^\infty_{lmN}|^2}{4\pi\omega_{mN}^2}\;,
\label{eq:EdotHres2}
\\
\left\langle\frac{dL_z^H}{dt}\right\rangle &=&
\sum_{lmN}\alpha_{lmN}
\frac{m|{\cal Z}^\infty_{lmN}|^2}{4\pi\omega_{mN}^3}\;.
\label{eq:LzdotHres2}
\end{eqnarray}
These formulas reproduce the flux-derived results given in
Sec.\ {\ref{sec:resI}}.

\bibliographystyle{apsrev}

\end{document}